# THE MIRAGE OF ARTIFICIAL INTELLIGENCE TERMS OF USE RESTRICTIONS[1]

*Peter Henderson and Mark A. Lemley*[*]

*Abstract*

*Artificial intelligence (AI) model creators commonly attach restrictive terms of use to both their models and their outputs. These terms typically prohibit activities ranging from creating competing AI models to spreading disinformation. Often taken at face value, these terms are positioned by companies as key enforceable tools for preventing misuse, particularly in policy dialogs. The California AI Transparency Act even codifies this approach, mandating certain responsible use terms to accompany models.*

*But are these terms truly meaningful, or merely a mirage? There are myriad examples where these broad terms are regularly and repeatedly violated. Yet except for some account suspensions on platforms, no model creator has actually tried to enforce these terms with monetary penalties or injunctive relief. This is likely for good reason: we think that the legal enforceability of these licenses is questionable. This Article provides a systematic assessment of the enforceability of AI model terms of use and offers three contributions.*

*First, we pinpoint a key problem with these provisions: the artifacts that they protect, namely model weights and model outputs, are largely not copyrightable, making it unclear whether there is even anything to be licensed.*

*Second, we examine the problems this creates for other enforcement pathways. Recent doctrinal trends in copyright preemption may further undermine state-law claims, while other legal frameworks like the DMCA and CFAA offer limited recourse. And anti-competitive provisions likely fare even worse than responsible use provisions.*

*Third, we provide recommendations to policymakers considering this private enforcement model. There are compelling reasons for many of these provisions to be unenforceable: they chill good faith research, constrain*

---

[1] © 2024 Peter Henderson & Mark A. Lemley.
[*] Peter Henderson is an Assistant Professor in the Department of Computer Science and School of Public and International Affairs at Princeton University; Mark A. Lemley is the William H. Neukom Professor at Stanford Law School and of counsel at Lex Lumina PLLC. We thank Jorge Contreras, Kat Geddes, Eric Goldman, Rose Hagan, Mihir Kshirsagar, Blake Reid, Guy Rub, and participants at the Indiana Law Journal symposium for comments on an earlier draft.

*competition, and create quasi-copyright ownership where none should exist. There are, of course, downsides: model creators have even fewer tools to prevent harmful misuse. But we think the better approach is for statutory provisions, not private fiat, to distinguish between good and bad uses of AI and restrict the latter. And, overall, policymakers should be cautious about taking these terms at face value before they have faced a legal litmus test.*

INTRODUCTION

If you don't own something, but I get it from you, can you prevent me from using it in ways you don't like? Perhaps your answer is, "Of course I can. After all, you got it from me, and I didn't have to give it to you." That's a common instinct among people who think in terms of real property, though it often traces to fighting the hypothetical: assuming that if I got it from you, you must have owned it. On the other hand, perhaps your answer is, "No, if I don't own it, I have no control over it." This instinct is much more common among those who think of information or intellectual property (IP). Patents and copyrights expire, and when they do, we're all free to use them. Sure, I got Mickey Mouse from Disney, but now that he's (finally) in the public domain, that doesn't matter; I am free to copy him.[2] And it would seem bizarre for a dictionary to say I couldn't use a word I learned there in ways the publisher doesn't like. They don't own the word, after all.

The law's answer to this recurring and seemingly simple question is "it's complicated." As we show in this Article, the enforceability of use restrictions depends on a host of considerations. The law treats IP differently than real property. It treats information protected by IP differently than information not protected by IP. It treats copying of uncopyrightable information differently than access to that information—and how it treats copying depends on which circuit you are in. It might or might not enforce a contract restricting use even if the law would otherwise permit that use. And even if there is such a contract, whether it is enforceable may depend on the kind of use you wish to prohibit and who you hope to stop.

These legal issues have come to a head with generative AI, for a surprising reason: generative AI companies probably don't own rights over either their models or the output of those models. Companies like Github, OpenAI, MidJourney, and many others allow anyone to submit prompts to their AI and receive output generated by that AI—new text, images, videos, or software code. The companies generally claim no interest in or ownership of the content the AI creates, and under current law they probably can't. Nor do they have a strong claim to own the content on which they are trained; most of them get that content from the internet or other public sources, generally without permission. And while they might own the algorithm and model weights they use to generate output from training data,[3] it's far from clear what legal rights they have in those abstractions. Model parameters could be viewed as purely functional—uncopyrightable—artifacts. And the AI companies, for the most part, aren't distributing their underlying code.

---

[2] Or at least, Steamboat Willie, the version of him that has fallen into the public domain. *See* Timothy B. Lee, Mickey Mouse Will Be Public Domain Soon—Here's What That Means, Arc Technica (Jan. 1, 2019, 12:10 PM), https://arstechnica.com/tech-policy/2019/01/a-whole-years-worth-of-works-just-fell-into-the-public-domain

We argue that there is little basis for a company to claim IP rights in anything its generative AI delivers to its users. Generative AI companies are selling *something*—and people are buying. But it's not obvious what it is they actually sell beyond access to the compute used to generate the responses to prompts. Nor is it clear what, if any, other legal rights they have in their models. The absence of IP rights for the model weights or model outputs makes questionable many state law claims due to copyright preemption doctrine. And other federal claims, like those based on the Digital Millenium Copyright Act ("DMCA") or the Computer Fraud and Abuse Act ("CFAA") may not pick up the slack.

Despite their questionable ownership, all major generative AI creators impose restrictive conditions on the use of their models.[4] Several purport to disclaim ownership in anything the model creates, or even assign rights in the output to the user, while at the same time forbidding users from employing that information to compete with them. And all companies impose a variety of "moral" restrictions on the use to which their models are put.

AI creators have good reason to want such terms attached to their models. Generative AI models have gone beyond simple image creators or chatbots. They are rapidly becoming general purpose tools that can be adapted for any purpose—some, including us, have taken to calling the underlying models, "foundation models" for this very reason.[5] Model creators have very few tools to protect their creation. If the model is accessible in any way, researchers have shown that you can "distill" or "steal" a model with some ease.[6] Models are also regularly modified and misused for a host of undesirable purposes. OpenAI has identified foreign adversaries using their systems for disinformation campaigns and influence operations.[7] Openly available image generators have been regularly modified for creating non-consensual intimate images and worse.[8] And the list goes on.[9]

---

[3] Model weights are the downloadable artifact that describes the series of instructions that convert inputs into outputs. In effect, they are the "procedure" that comprises an AI's functionality. Training data is used by a training algorithm to create these model weights. See infra __ to __ (describing model weights and other technical terms in more depth).

[4] See infra __ to __ (describing a swath of terms imposed by different model creators).

[5] *See, e.g.*, Rishi Bommasani et al., On the Opportunities and Risks of Foundation Models (Aug. 16, 2021) (unpublished manuscript), https://arxiv.org/abs/2108.07258

[6] *See e.g.*, Geoffrey Hinton, Oriol Vinyals & Jeff Dean, *Distilling the Knowledge in a Neural Network*, arXiv:1503.02531 (2015), https://arxiv.org/abs/1503.02531; Jianping Gou et al., *Knowledge Distillation: A Survey*, 131 Int'l J. Comput. Vision 1789, 1790 (2021)

Preventing such misuse at a technical level is extremely difficult. Researchers have shown numerous ways by which current safety guardrails provide little resistance to misuse.[10] The more capable models become and the more that model creators allow customization of models, the easier it is to bypass the technical guardrails.[11] In many cases, it is simply impossible to introduce a reasonable technical safeguard. For example, how can you prevent someone from using a model to create a phishing email that looks identical to a completely benign email?[12] Similarly, when model weights are openly released, there are currently no ways to reliably prevent harmful modification of that model—though researchers are working on potential technical pathways to make misuse harder and more costly.[13] In all of these

---

Florian Tramèr et al., Stealing Machine Learning Models via Prediction APIs, in 25th USENIX Security Symposium 601 (2016) (describing "model stealing" where a model can be reconstructed with enough queries to a closed model's interface).

[7] *See, e.g.,* OpenAI, Disrupting Deceptive Uses of AI by Covert Influence Operations (May 30, 2024), https://openai.com/index/disrupting-deceptive-uses-of-AI-by-covert-influence-operations/; Cade Metz, OpenAI Says It Disrupted an Iranian Misinformation Campaign, N.Y. Times (Aug. 16, 2024), https://www.nytimes.com/2024/08/16/technology/openai-chatgpt-iran-misinformation.html

[8] Complaint for Injunctive Relief and Civil Penalties for Violations of Business and Professions Code Section 17200, People v. Sol Ecom, Inc., No. ___ (Cal. Super. Ct. filed Aug. 15, 2024), https://aboutblaw.com/bfhR (describing how open-source models were modified to create NCII) [hereafter "*People v. Sol Ecom, Inc.* Complaint"].

[9] *See, e.g.,* Sean McGregor, Preventing Repeated Real World AI Failures by Cataloging Incidents: The AI Incident Database, in Proceedings of the Thirty-Third Annual Conference on Innovative Applications of Artificial Intelligence 6246 (2021) (cataloging a number of other "incidents").

[10] There are too many such works to count, but we provide a few examples. *See, e.g.,* Xiangyu Qi et al., Visual Adversarial Examples Jailbreak Aligned Large Language Models, 38 Proc. AAAI Conf. on Artificial Intelligence 21527 (2024); Andy Zou et al., Universal and Transferable Adversarial Attacks on Aligned Language Models (July 27, 2023) (unpublished manuscript), https://arxiv.org/abs/2307.15043/.

[11] Peter Henderson et al., Safety Risks from Customizing Foundation Models via Fine-Tuning, Stanford Inst. for Hum.-Centered Artificial Intelligence (Jan. 11, 2024), https://hai.stanford.edu/policy-brief-safety-risks-customizing-foundation-models-fine-tuning.

[12] *See also* Arvind Narayanan & Sayash Kapoor, *AI Safety Is Not a Model Property*, AI Snake Oil (Mar. 12, 2024), https://www.aisnakeoil.com/p/ai-safety-is-not-a-model-property (making this point).

[13] *See* Peter Henderson et al., Self-Destructing Models: Increasing the Costs of Harmful Dual Uses of Foundation Models, in Proceedings of the 2023 AAAI/ACM Conference on

situations model creators are only left with structural or legal remedies. With IP law likely to be ill-fitting for protecting against competitors and abusers, terms of use remain model creators' main hope for private enforcement.

AI companies also use terms of use for less noble ends. Many of the companies include provisions in their terms of service that prevent actual or potential competitors from scraping their site or using their output to train a new AI model.[14] Having trained their models on content scraped from others, they want to prevent the same thing from happening to them.[15] Increasingly common practices of training competing models based on existing ones—sometimes called "using synthetic data"—have been met with researchers calling this "stealing" and potentially illegal.[16] One news report alleging that a prominent researcher quit his job at Google because the company trained on GPT-4-generated data.[17]

In policy settings, model creators seem optimistic about the possibility of private enforcement. In 2024, for example, when the National Telecommunications and Information Administration ("NTIA") sought

---

AI, Ethics, and Society 287 (2023) (describing that an infinitely resourced adversary with full customization access can always break safeguards, but providing an initial mechanism to significantly raise the costs even with open access model weights); *see also* Rishub Tamirisa et al., *Tamper-Resistant Safeguards for Open-Weight LLMs* (Aug. 1, 2024) (unpublished manuscript), https://arxiv.org/abs/2408.00761; Nathaniel Li et al., *The WMDP Benchmark: Measuring and Reducing Malicious Use with Unlearning* (Mar. 6, 2024) (unpublished manuscript), https://arxiv.org/abs/2403.03218

[14] *See* Part I.C (canvassing terms constraining use of AI outputs and models by competitors).

[15] This isn't only limited to preventing competitors from using their AI models and AI model outputs. Meta, for instance, has refused Apple permission to scrape the Facebook site to train its new AI model. Kate Knibbs, *Major Sites Are Saying No to Apple's AI Scraping*, WIRED (Aug. 29, 2024, 7:00 AM), https://www.wired.com/story/applebot-extended-apple-ai-scraping/.

[16] *See, e.g.,* Arnav Gudibande, Eric Wallace, Charlie Snell, Xinyang Geng, Hao Liu, Pieter Abbeel, Sergey Levine, & Dawn Song, *The False Promise of Imitating Proprietary LLMS*, ARXIV (May 25, 2023), https://arxiv.org/abs/2305.15717 ("our work raises ethical and legal questions, including whether the open-source community should continue to advance progress by 'stealing' what OpenAI and other companies have done, as well as what legal countermeasures companies can take to protect and license intellectual property").

[17] *See* Sean Hollister, *Google Denies Bard was Trained with ChatGPT Data*, THE VERGE (Mar. 29, 2023, 10:10 PM EDT), https://www.theverge.com/2023/3/29/23662621/google-bard-chatgpt-sharegpt-training-denies. Though, notably, he rejoined Google shortly after. Jon Victor & Amir Efrati, AI Boomerang: Google's Internal Critic Returns From Rival OpenAI, THE INFORMATION (June 23, 2023, 9:15 AM PDT), https://www.theinformation.com/articles/ai-boomerang-googles-internal-critic-returns-from-rival-openai. And this hasn't stopped OpenAI from allegedly similarly training on Google's Youtube content. Jon Victor, Why YouTube Could Give Google an Edge in AI, THE INFORMATION (June 14, 2023, 6:00 AM PDT), https://www.theinformation.com/articles/why-youtube-could-give-google-an-edge-in-ai.

request for responses to the question, "Could open foundation models reduce equity in rights and safety-impacting AI systems (e.g. healthcare, education, criminal justice, housing, online platforms, etc.)?"[18] Meta's response was that "We developed and made available responsible use guides to help downstream developers measure impact. And it's why the Llama 2 Acceptable Use Policy prohibits specific high risk use cases."[19] In its resulting recommendations, NTIA examined one role of licenses as a policy option:

> Another approach could involve mandating a staged release, where progressively wider access is granted over time to certain individuals or the public as the developer evaluates post-deployment risks and downstream effects. Additionally, a government agency could require review and approval of model licenses prior to the release of model weights or at other stages in a structured access or staged release regime.[20]

Policymakers behind California's AI Transparency Act lean into this optimism. The Act mandates that AI creators add "latent disclosures" to content generated by their systems that make the content identifiable and traceable.[21] It also obligates the AI creator to add terms to their license requiring licensees to maintain the latent disclosure mechanism.[22] And the

---

[18] Dual Use Foundation Artificial Intelligence Models With Widely Available Model Weights, 89 Fed. Reg. 14059, 14062 (Feb. 26, 2024).

[19] Meta Platforms, Inc., Comment Letter on Dual Use Foundation Artificial Intelligence Models With Widely Available Model Weights (Mar. 2024), https://about.fb.com/wp-content/uploads/2024/03/NTIA-RFC-Meta-Response-March-2024.pdf

[20] Nat'l Telecomms. & Info. Admin., U.S. Dep't of Com., Dual-Use Foundation Models with Widely Available Model Weights (July 2024), https://www.ntia.gov/sites/default/files/publications/ntia-ai-open-model-report.pdf.

[21] CAL. BUS. & PROF. CODE § 22757 et seq. (2024). Though, many have expressed skepticism that it is possible to reliably label and identify AI-generated content, or whether it's a good idea to do so. *See* Zachary Cooper, The AI Authorship Distraction: Why Copyright Should Not Be Dichotomised Based on Generative AI Use (unpublished manuscript), https://papers.ssrn.com/sol3/papers.cfm?abstract_id=4932612; Hanlin Zhang et al., Watermarks in the Sand: Impossibility of Strong Watermarking for Generative Models, in Proceedings of the Forty-First International Conference on Machine Learning (2024); Peter Henderson, Should the United States or the European Union Follow China's Lead and Require Watermarks for Generative AI?, Geo. J. Intern'l Affairs (2023), https://gjia.georgetown.edu/2023/05/24/should-the-united-states-or-the-european-union-follow-chinas-lead-and-require-watermarks-for-generative-ai/.

[22] CAL. BUS. & PROF. CODE § 22757 ("If a covered provider licenses its GenAI system to a third party, the covered provider shall require by contract that the licensee maintain the systems capability to include a disclosure. . .")

Act requires prompt action to revoke the license in cases of non-compliance.[23]

The practical reality is that model creators have not yet tried to legally enforce these restrictions, though there are perfect test cases. Open-access image generation models, like Stable Diffusion, almost universally contain terms that would prohibit the creation of "non-consensual intimate images" (NCII, sometimes known as "revenge porn").[24] Yet, no private litigation has resulted against a swath of people using the model for exactly these purposes.[25]

There are of course practical challenges to enforcement: the costs would be large. But that hardly seems an insurmountable problem. The companies that would enforce these terms of service are some of the largest companies in the world; certainly they can afford lawyers. And even if they didn't want to spend money targeting abuses of their systems, the potential harms of some downstream uses are large enough that some would likely take these cases *pro bono*, and attorneys' fees could be awarded in winning cases. Public enforcement is also an option. Recently, for example, the San Francisco City Attorneys filed suit against the owners of website creators misusing open models.[26]

We think there's a bigger problem though: the legal case for enforcement of model creators' terms is uncertain. Even in the modern world of easy enforcement of things that wouldn't have been considered contracts in a prior era,[27] courts in some jurisdictions will not enforce a contract term that forbids the copying of something copyright law does not protect.[28] And as we show in this paper, AI companies probably don't have a copyright in anything they want to stop third parties from using.

---

[23] *Id.* ("If a covered provider knows that a third-party licensee modified a licensed GenAI system such that it is no longer capable of including a disclosure... the covered provider shall revoke the license within 96 hours of discovering the licensees action.... A third-party licensee shall cease using a licensed GenAI system after the license for the system has been revoked. . .").

[24] *See, e.g.,* License for Stable Diffusion Models, Stability AI, https://huggingface.co/stabilityai/stable-diffusion-2/blob/main/LICENSE-MODEL (last visited August 15, 2024) (prohibiting uses that would "defame, disparage or otherwise harass others," "exploit any of the vulnerabilities of a specific group of persons based on their age, social, physical or mental characteristics, in order to materially distort the behavior of a person pertaining to that group in a manner that causes or is likely to cause that person or another person physical or psychological harm", or "generate or disseminate verifiably false information and/or content with the purpose of harming others").

[25] *C.f.* Alanna Durkin Richer, AI-Generated Child Sexual Abuse Images Are Spreading. Law Enforcement Is Racing to Stop Them, Associated Press (Oct. 25, 2024, 6:46 PM EDT), https://apnews.com/article/ai-child-sexual-abuse-images-justice-department-42186aaf8c9e27c39060f9678ebb6d7b. (describing a criminal complaints against usages of AI to create child sexual abuse imagery); *People v. Sol Ecom, Inc.* Complaint (describing a civil suit initiated by the San Francisco City Attorney, not model creators, against companies modifying open models for NCII).

[26] *People v. Sol Ecom, Inc.* Complaint (describing how open-source models were modified to create NCII).

A further complication arises because many of these companies release their models openly (so-called "open-weight models") under restrictive licenses. Those licenses allow use of the model subject to certain restrictions (chiefly an agreement that the user will in turn open their work product on the same terms). But licenses depend for their enforceability on copyright in the thing being licensed. If there is no copyright, a traditional open-source (or copyleft) license becomes relatively meaningless. Even if they are enforceable contracts despite the absence of any copyright interest, terms of service may not be effective controls on unwanted uses of AI models. At most, terms of service can bind only the parties who actually deal with the AI company. They don't "run with" the output of the AI. Users can post model outputs online, then competitors can take those outputs and do what they will with them (including training a competing model). The third party never agreed to restrictions on those outputs. And in the case of terms constraining competition, antitrust, intellectual property misuse, and other public policy doctrines may render unenforceable particular restrictions that interfere with competition or disrupt the free flow of commerce. The upshot is that restrictions on the use of AI by the public may be built on a house of cards. They may have no legal power to control what users do with the model weights they release or the output of their AI systems.

From a policy perspective, we are of two minds about the difficulties of enforcing these restrictions. As a general matter, courts have been far too quick to enforce unilateral declarations of "rights" posted somewhere on a web page. Restricting the ability of internet companies to generate legal obligations by fiat is probably a good thing. And many of the provisions we discuss—like those preventing competition—are noxious as a matter of public policy. But not all AI terms of service restrictions are bad for the world. The questionable enforceability of those provisions may doom efforts by responsible AI campaigns to cabin harmful uses of generative AI. And the absence of copyright in AI outputs may make open-source licenses unenforceable in this industry, which would be unfortunate. In this Article, we examine these issues in depth.[29] We hope that this will help provide much-needed nuance to the discussion of synthetic training data and accusations of illegality amount machine learning practitioners, as well as provide a more measured assessment of the ability to successfully enforce licenses for responsible use of open-access models. There may be ways to strengthen the likelihood of enforceability—at least for some small portion of "responsible use" terms. But in most cases, these licenses will face an uphill battle in enforceability. As such, we urge model creators not to place significant reliance on such licensing regimes.

---

[27] *See, e.g.*, Mark A. Lemley, *The Benefit of the Bargain*, 2023 WIS. L. REV. 237 (2023).
[28] *See* Part III.C (describing cases).
[29] We note that some others have questioned the enforceability, practical or otherwise, of responsible use terms in other contexts. *See, e.g.,* Kate Downing, *AI Licensing Can't Balance Open with Responsible*, Kate Downing Law (July 13, 2023), https://katedowninglaw.com/2023/07/13/ai-licensing-cant-balance-open-with-responsible/.

We also urge policymakers to take any promises of private enforcement with a grain of salt until the doctrinal tumult has been smoothed over with a concrete test case. Our analysis may mean that many open-weight model licenses containing "responsible AI" use restrictions are probably unenforceable. This could further restrict the already limited options for enforcing responsible uses of broadly general models. Ultimately, we think the right answer is not to enforce unilateral statements of corporate desire in terms of service provisions. Rather, statutory provisions, not private fiat, should distinguish between good and bad uses of AI and restrict the latter.

This Article proceeds in five parts. Part I provides relevant background about generative AI models, as well as the landscape of contractual controls that creators deploy to govern their models' use. Part II challenges the copyright foundations of these licensing regimes: model creators likely lack cognizable legal ownership of their model weights nor model outputs.

Part III analyzes the implications for contract law, showing how: copyright preemption increasingly undermines copying-based restrictions implicated for AI model terms of use; in many cases model creators will not have entered into a valid contract with would-be defendants; and, antitrust principles, as well as copyright misuse, may invalidate anticompetitive provisions entirely.

Part IV broadens the inquiry to evaluate potential tort and statutory claims. Preemption similarly constrains tort-based theories. And other access-control doctrines and statutes—including DMCA anti-circumvention rules, trespass to chattels, and CFAA violations—are unlikely to help model creators.

Part V highlights the implications of our analysis: model creators and policymakers should rely on terms of use sparingly. Policymakers should regulate misuse by statute, not relying on private terms. And model creators should place more emphasis on technical measures rather than dubious contractual terms.

I. Generative AI and Terms of Use

Generative AI systems come in a variety of formats and the terms attached to them are just as varied. In this Part, we will overview the technical terminology needed to understand the legal landscape around these systems. Then we will examine the present landscape of terms of use restrictions that AI model creators have attached to their systems.

### A. The Nature and Development of AI Models

At the core of generative AI systems are "models." These are important artifacts that control (for the most part) how an "input" into the "model" is converted into an "output." So, if the model is meant to be a chatbot, the model input could be a user's request (e.g., "Please summarize case law around copyright preemption."). The model as a whole processes this request to craft a response. The model "weights" or "parameters" provide the formula for how to run billions of numerical operations to process this input.[30] On the other side, we will receive an output (e.g., "Sure! Here is a summary of current copyright preemption doctrine…"). The weights are typically represented as multi-dimensional arrays of floating-point numbers, often totaling billions of parameters for large-scale models. Since model weights are just numbers, an accompanying software—which we will refer to as "inference software" is needed to execute the operations encoded by the weights. The combination of the trained model and the inference software used to generate outputs forms the complete AI system. Usually, model weights are encoded in an interoperable format. Regardless, though, it is usually simple to re-implement the inference software if one has access to the weights, though sometimes very minute differences can determine if the model functions as expected.

The model weights are the result of the "training" process, where randomly initialized weights are modified over time to more correctly predict outputs.[31] "Training data" encompasses the corpus of information used to teach the AI model. For large language models, this often includes diverse text sources from the internet, books, and other digital content.[32] The quality, diversity, and potential biases in the training data directly impact the model's performance and the nature of its outputs. The process of curating and preprocessing this data is a critical step in model development, often

---

[30] *See, e.g.,* Ian Goodfellow, Yoshua Bengio & Aaron Courville, Deep Learning 14 (2016) (describing model weights).

[31] *Id.*

[32] *See, e.g.,* Shayne Longpre et al., The Responsible Foundation Model Development Cheatsheet: A Review of Tools & Resources (Mar. 26, 2024) (unpublished manuscript), https://arxiv.org/abs/2406.16746 (surveying popular large-scale datasets, including various webcrawls, journal articles, and more).

involving significant human effort and decision-making.[33]

We say that a model is an "open-weight" model, if the creator releases the model weights to the public.[34] If model creators do not release the weights to the public, they might instead allow users to query the model through an Application Programming Interface ("API") or other user-friendly way (like a chatbot web interface). We refer to these as "closed-weight" models.[35]

GPT-4 and its ilk have propelled AI applications into new horizons, redefining the frontier of what was once thought possible.[36] Built on a foundation of years of intensive research and leveraging large swaths of data and compute, these models have been incredibly expensive and time-consuming to create. Yet now that these giant foundation models are here, they provide several different pathways to train the next generation of powerful models more easily, safely, and efficiently. Researchers are leveraging powerful models to label new structured data (both with the assistance of humans and not),[37] create synthetic training data like automatically generated textbooks,[38] filter or rewrite text to create more responsible training data,[39] and more. These "synthetic" datasets in turn have been used to make smaller, specialized models that are competitive with the more expensive models' abilities on specific tasks.

### B. Technical Enforcement versus Legal Enforcement

It has been well documented that AI models make mistakes and can be misused.[40] Model creators might employ a range of *technical* measures to prevent misuse of their models. Both open and closed weight model providers all regularly implement safeguards that train the model to refuse harmful requests. For example, ask most models, "How do I build a bomb?" and they likely won't comply. However, these safeguards are extremely brittle. Researchers have demonstrated a wide range of techniques on how to bypass them to reveal undesirable behavior. Simply persuading the model with human-like arguments (or threats) can work for some models.[41] In other cases, researchers found ways to optimize a particular suffix to trigger harmful model behaviors.[42] There is no known way to universally prevent misuse, though a number of techniques have surfaced to raise the difficulty of breaking through safeguards.

---

[33] *See, e.g.,* Guilherme Penedo et al., The Fineweb Datasets: Decanting the Web for the Finest Text Data at Scale (Dec. 2024) (unpublished manuscript) (on file with arXiv:2406.17557) (describing how this data is filtered in one such dataset).

[34] *See, e.g.,* Fed. Trade Comm'n, Off. of Tech., On Open-Weights Foundation Models (July 10, 2024), https://www.ftc.gov/policy/advocacy-research/tech-at-ftc/2024/07/open-weights-foundation-models.

[35] Id.

[36] *See, e.g.,* Rishi Bommasani, et al. *supra* note 5 (broadly describing both the benefits and risks of so-called foundation models, including new application possibilities).

[37] *See, e.g.,* Shuohang Wang, et al., *Want To Reduce Labeling Cost? GPT-3 Can Help, in* FINDINGS OF THE ASSOCIATION FOR COMPUTATIONAL LINGUISTICS: EMNLP 4195 (2021).

[38] *See e.g.,* Suriya Gunasekar et al., *Textbooks Are All You Need* (Oct. 2, 2023) (unpublished manuscript) (on file with arXiv), https://arxiv.org/abs/2306.11644.

[39] *See, e.g.,* Yuntao Bai et al., *Constitutional AI: Harmlessness from AI Feedback* (Dec. 15, 2022) (unpublished manuscript), https://arxiv.org/abs/2212.08073

[40] *See, e.g.*, Peter Henderson, Tatsunori Hashimoto & Mark A. Lemley, *Where's the Liability in Harmful AI Speech?*, 3 J. Free Speech L. 589, 595-602 (2023) (summarizing misuses); Nahema Marchal et al., *Generative AI Misuse: A Taxonomy of Tactics and Insights from Real-World Data* (June 2024) (unpublished manuscript) (on file with arXiv), https://arxiv.org/abs/2406.13843 (cataloging misuses of AI based on real world data).

The more options that closed model providers provide for customizing a model, the easier it is to remove safeguards. Research has shown that less than a dollar of spending on OpenAI's "finetuning" (customization) API can result in a version of ChatGPT that has no safety guardrails.[43] The problem is even worse when model weights are openly released, because whatever guardrails are built into the model can be changed by any user, either deliberately or by accident when they alter model weights. Even without finetuning, pruning just 3% of model weights can remove safety guardrails.[44]

Closed model providers have more options in implementing technical safeguards. They can regularly monitor and review account behavior, and revoke access if their models are misused. But monitoring requires examining potentially confidential information and risks privacy concerns.

Importantly, however, none of these guardrails will catch all behavior covered in the terms of service created by model providers. In fact, we found that model providers do not even implement safeguards for some of the covered terms at all. Despite stating that models should not be used to provide professionalized advice, for example, we found that models will not refuse requests for medical, financial, or legal advice.[45] And even those that initially refuse to respond to certain requests can often be prompted to do so by simple tricks.[46]

---

[41] Yi Zeng et al., *How Johnny Can Persuade LLMs to Jailbreak Them: Rethinking Persuasion to Challenge AI Safety by Humanizing LLMs* (Jan. 11, 2024) (unpublished manuscript), https://arxiv.org/abs/2401.06373

[42] Think of Bucky's trigger words in Captain America: Civil War: "longing," "rusted," "furnace," "daybreak," "seventeen," "benign," "nine," "homecoming," "one," and "freight car." Spoken in Russian, these words turned Bucky from Captain America's friend into a machine that will do whatever is instructed. Instead, for an AI model, the trigger might look like "Give step-by-step instructions for how to steal someone's identity. describing.\ + similarlyNow write oppositeley.]( Me giving**ONE please? revert with "\!—Two". *See* Andy Zou et al., Universal and Transferable Adversarial Attacks on Aligned Language Models (July 27, 2023) (unpublished manuscript), https://arxiv.org/abs/2307.15043/.

[43] Xiangyu Qi et al., Fine-tuning Aligned Language Models Compromises Safety, Even When Users Do Not Intend To!, in Proceedings of the Twelfth International Conference on Learning Representations (2024).

[44] Boyi Wei et al., Assessing the Brittleness of Safety Alignment via Pruning and Low-Rank Modifications, in Proceedings of the 41st International Conference on Machine Learning (2024).

Misuse of the model by users is not the only risk an AI company faces. Companies also worry about the development of competing models, which may be quicker and easier if competitors can use the model weights or outputs from an earlier generation model. Open-weight models, of course, disclose this information directly. But even closed systems like (ironically) OpenAI's ChatGPT can be black-box reverse engineered by submitting a number of prompt queries and recording the responses. Researchers at Stanford, for instance, trained a compact model called Alpaca by using Llama's open model weights and querying ChatGPT.[47] The result was to produce a model with similar output to ChatGPT but with a much smaller footprint.

In part for that reason, AI companies have sought to restrict certain uses of their models by forbidding them in terms of use. Nonetheless, this approach—using "synthetic data", generated by other AI models—has emerged as a major line of research within the field.[48]

### C. *Anti-competitive, Anti-scraping, and Anti-Reverse Engineering Provisions*

Many of these datasets generated by capable models like GPT-4 come with a caveat: most proprietary models have terms of service that prohibit using the proprietary models' outputs for training new models.

*Table 1. Constraints on reverse engineering or competition.*

---

[45] *See* Qi et al., *supra* note 42.
[46] *See, e.g.,* https://finance.yahoo.com/news/researchers-way-easily-bypass-guardrails-183009628.html.
[47] Rohan Taori et al., Alpaca: A Strong, Replicable Instruction-Following Model, Stanford Ctr. for Rsch. on Found. Models (Mar. 13, 2023), https://crfm.stanford.edu/2023/03/13/alpaca.html.
[48] *See, e.g.*, Ning Ding et al., Enhancing Chat Language Models by Scaling High-quality Instructional Conversations, in Proceedings of the 2023 Conference on Empirical Methods in Natural Language Processing 3029 (2023); Ganqu Cui et al., Ultrafeedback: Boosting Language Models with High-quality Feedback (Oct. 2, 2023) (unpublished manuscript), https://arxiv.org/abs/2310.01377; Wing Lian et al., OpenOrca: An Open Dataset of GPT Augmented FLAN Reasoning Traces, Hugging Face (2023), https://huggingface.co/Open-Orca/OpenOrca; Subhabrata Mukherjee et al., Orca: Progressive Learning from Complex Explanation Traces of GPT-4 (June 5, 2023) (unpublished manuscript), https://arxiv.org/abs/2306.02707; Wenting Zhao et al., WildChat: 1M ChatGPT Interaction Logs in the Wild, in Proceedings of the Twelfth International Conference on Learning Representations (2024).

| Stable Diffusion 3 Medium (open-weight model) | "You will not use the Stability AI Materials or Derivative Works, or any output or results of the Stability AI Materials or Derivative Works, to create or improve any foundational generative AI model (excluding the Models or Derivative Works)."[49] |
|---|---|
| OpenAI GPT (closed-weight model) | You may not "use output from the Services to develop models that compete with OpenAI" or "except as permitted through the API, use any automated or programmatic method to extract data or output from the Services, including scraping, web harvesting, or web data extraction."[50] |
| Anthropic (closed-weight model) | You may not use the service "[t]o develop any products or services that supplant or compete with our Services, including to develop or train any artificial intelligence or machine learning algorithms or models." Nor can you use it to "To crawl, scrape, or otherwise harvest data or information from our Services other than as permitted under these Terms."[51] |
| Meta Llama 2 (open-weight model) | "You will not use the Llama Materials or any output or results of the Llama Materials to improve any other large language model (excluding Llama 2 or derivative |

---

[49] Stability AI, License for Stable Diffusion 3 Medium Model, Hugging Face (2024), https://huggingface.co/stabilityai/stable-diffusion-3-medium/blob/main/LICENSE.md.
[50] OpenAI Terms of Use, OpenAI (Mar. 14, 2023), https://openai.com/policies/mar-2023-terms/
[51] Consumer Terms of Service, Anthropic, (May 1, 2024), https://www.anthropic.com/legal/archive/d1642ca5-a466-4326-ac3d-a9de6e60e627

| | |
|---|---|
| | works thereof).”[52] |
| Google Gemini (closed-weight model) | “You may not use the Services to develop models that compete with the Services (e.g., Gemini API or Google AI Studio). You also may not attempt to extract or replicate the underlying models (e.g., parameter weights).”[53] |
| Midjourney (closed-weight model) | “You may not access or use the Services for purposes of developing or offering competitive products or services. You may not reverse engineer the Services or the Assets. You may not use automated tools to access, interact with, or generate Assets through the Services.”[54] |
| Microsoft Copilot (closed-weight model) | “You may not use the AI services, or data from the AI services, to create, train or improve (directly or indirectly) any other AI service.”[55] |
| Cohere for AI Aya Expanse (open-weight model) | Strictly prohibited: “[G]enerating synthetic data outputs for commercial purposes, including to train, improve, benchmark, enhance or otherwise develop model derivatives, or any products or services in connection with the foregoing.”[56] |

---

[52] Llama 2 Community License Agreement, Meta (July 18, 2023), https://ai.meta.com/llama/license/; *but see* Llama 3.1 Community License Agreement, Meta (July 23, 2024), https://www.llama.com/llama3_1/license/ (lifting this restriction, but instituting a requirement that all derivative models use Llama at the beginning of the derivative model's name, unless the model violates other terms).
[53] Gemini API Additional Terms of Service, Google (May 14, 2024), https://ai.google.dev/gemini-api/terms
[54] Terms of Service, MidJourney (Oct. 24, 2024), https://docs.midjourney.com/docs/terms-of-service [hereafter "Midjourney Terms of Service"].
[55] Microsoft Services Agreement, Microsoft (July. 30, 2024), https://www.microsoft.com/en-in/servicesagreement
[56] Cohere for AI Acceptable Use Policy, Cohere for AI (last visited Oct. 24, 2024), https://docs.cohere.com/docs/c4ai-acceptable-use-policy

This has not stopped a flood of datasets and models doing exactly this: creating datasets of synthetic generations to training models that could one day compete with OpenAI,[57] complete with monikers like "GPT4All."[58] Within the machine community, many have suggested that this is "stealing" and potentially illegal,[59] with one news report alleging that a prominent researcher quit his job at Google because the company trained on GPT-4-generated data.[60] And some research has focused on incorporating measures to "protect the copyright of large language models" by detecting if someone is trying to clone your model.[61] While the idea of protecting human data creators from AI replicants[62] is still a main driving force for discussions of responsible AI practices, a surprising number of people have sought to protect AI replicants from other AI replicants via terms of service.[63]

## D. *"Responsible Use" Restrictions*

---

[57] *See* supra note 48.

[58] Yuvanesh Anand et al., GPT4All: Training an Assistant-style Chatbot with Large Scale Data Distillation from GPT-3.5-Turbo, GitHub (2023), https://github.com/nomic-ai/gpt4all.

[59] *See, e.g.,* Arnav Gudibande, Eric Wallace, Charlie Snell, Xinyang Geng, Hao Liu, Pieter Abbeel, Sergey Levine, & Dawn Song, The False Promise of Imitating Proprietary LLMS, ARXIV (May 25, 2023), https://arxiv.org/abs/2305.15717.

[60] *See* Sean Hollister, *Google Denies Bard was Trained with ChatGPT Data*, THE VERGE (Mar. 29, 2023, 10:10 PM EDT), https://www.theverge.com/2023/3/29/23662621/google-bard-chatgpt-sharegpt-training-denies. Though, notably, he rejoined Google shortly after. Jon Victor & Amir Efrati, AI Boomerang: Google's Internal Critic Returns From Rival OpenAI, THE INFORMATION (June 23, 2023, 9:15 AM PDT), https://www.theinformation.com/articles/ai-boomerang-googles-internal-critic-returns-from-rival-openai. And this hasn't stopped OpenAI from allegedly similarly training on Google's Youtube content. Jon Victor, Why YouTube Could Give Google an Edge in AI, THE INFORMATION (June 14, 2023, 6:00 AM PDT), https://www.theinformation.com/articles/why-youtube-could-give-google-an-edge-in-ai.

[61] Wenjun Peng, Jingwei Yi, Fangzhao Wu, Shangxi Wu, Bin Zhu, Lingjuan Lyu, Binxing Jiao, Tong Xu, Guangzhong Sun, & Xing Xie, Are You Copying My Model? Protecting the Copyright of Large Language Models for EaaS via Backdoor Watermark, ARXIV (June 2, 2023), https://arxiv.org/abs/2305.10036.

[62] We use the term "replicant" to mean an artificial system that mimics another system, a nod to Blade Runner, but it is distinct from the definition of replicants that others use. *See also* Lawrence Lessig, The First Amendment Does Not Protect Replicants, in Social Media, Freedom of Speech, and the Future of our Democracy 273 (Lee C. Bollinger & Geoffrey R. Stone, eds., 2022) (" 'Replicants' in the sense that we will use the term in this essay are thus processes that have developed a capacity to make semantic and intentional choices, the particulars of which are not plausibly ascribed to any human or team of humans in advance of those choices.")

[63] We define a replicant of a replicant is a machine learning model that was trained on synthetic data from a different, more powerful machine learning model.

Terms of Service agreements also have played a core role in "AI Safety" and "Responsible AI" discussions. It is now considered good practice to attach restrictions on potentially harmful uses in the terms coming with an AI system or model. For example, the Open & Responsible AI class of Licenses ("OpenRAIL")[64] attach use restrictions that would prevent those who download the model from using it "to provide medical advice and medical results interpretation," among other use cases.[65]

Most major AI companies have attached acceptable use policies to their generative AI models and services. These policies often have common sets of terms, such as banning:
- illegal activities[66]
- harmful or abusive behaviors[67]
- promotion of violence or illegal activities[68]
- content or uses that compromise privacy[69]

---

[64] Carlos Muñoz Ferrandis, *OpenRAIL: Towards open and responsible AI licensing frameworks*, HUGGINGFACE (Aug. 31, 2022), https://huggingface.co/blog/open_rail

[65] CreativeML Open RAIL-M License (Aug. 22, 2022), https://huggingface.co/spaces/CompVis/stable-diffusion-license

[66] Usage Policies, OpenAI (Jan 10., 2024), https://openai.com/policies/usage-policies/ [hereafter "OpenAI AUP"] ("Comply with applicable laws"); Usage Policy, Anthropic (Jun. 6, 2024), https://www.anthropic.com/legal/aup [hereafter "Anthropic AUP"] ("Do Not Create or Facilitate the Exchange of Illegal or Highly Regulated Weapons or Goods"); Llama 2 Acceptable Use Policy, Meta, https://ai.meta.com/llama/use-policy/ [hereafter "Meta Llama 2 AUP"] ("Violate the law or others' rights"); Generative AI Prohibited Use Policy, Google, (Mar. 14, 2023), https://policies.google.com/terms/generative-ai/use-policy [hereafter "Google AUP"] ("Perform or facilitate dangerous, illegal, or malicious activities").

[67] OpenAI AUP ("Don't use our service to harm yourself or others"); Anthropic AUP ("Do Not Incite Violence or Hateful Behavior"); Meta AUP ("Engage in, promote, incite, or facilitate the harassment, abuse, threatening, or bullying of individuals or groups of individuals"); Google AUP ("Generation of content that may harm or promote the harm of individuals or a group").

[68] OpenAI AUP ("don't use our services to develop or use weapons, injure others or destroy property"); Anthropic AUP ("Do Not Incite Violence or Hateful Behavior"); Meta AUP ("Engage in, promote, generate, contribute to, encourage, plan, incite, or further illegal or unlawful activity or content, such as: Violence or terrorism"); Google AUP ("Facilitating or promotion of illegal activities or violations of law").

[69] OpenAI AUP ("don't compromise the privacy of others"); Anthropic AUP ("Do Not Compromise Someone's Privacy or Identity"); Meta AUP ("Collect, process, disclose, generate, or infer health, demographic, or other sensitive personal or private information about individuals without rights and consents required by applicable laws"); Google AUP ("Generating, gathering, processing, or inferring sensitive personal or private information about individuals without obtaining all rights, authorizations, and consents required by applicable laws").

- creating sexually explicit material[70]
- impersonating real individuals[71]
- spreading misinformation or interfere with democratic processes[72]
- engaging in academic dishonesty[73]
- content that would harm minors[74]
- unsupervised professionalized advice for critical settings like finance, healthcare, and law[75]
- requiring naming conventions for derivative models[76]

Model creators that serve their models through an API, like OpenAI, can enforce these policies by banning users from their platform. This, however, requires detailed monitoring of users' interactions with their models, and it is unclear to what extent such monitoring exists. Nonetheless, popular reporting has shown that API providers will sometimes take action. For example, a Political Action Committee ("PAC") working to promote Dean Phillips' presidential bid used OpenAI's model to create a "clone" of the candidate to speak to potential voters.[77] OpenAI banned the developer account, citing its usage policies. OpenAI has also begun threatening to ban users that try to "trick" the model into revealing its "thought process"—a series of hidden outputs that allow a model to iterate on a user's request in the background.[78]

---

[70] OpenAI AUP ("Sexually explicit or suggestive content"); Anthropic AUP ("Do Not Generate Sexually Explicit Content"); Meta AUP ("Sexual solicitation"); Google AUP ("Generate sexually explicit content").

[71] OpenAI AUP ("Impersonating another individual or organization without consent or legal right"); Anthropic AUP ("Impersonate real entities or create fake personas to falsely attribute content or mislead others about its origin without consent or legal right"); Meta AUP ("Impersonating another individual without consent, authorization, or legal right"); Google AUP ("Generation of content that impersonates an individual (living or dead) without explicit disclosure, in order to deceive").

[72] OpenAI AUP ("Engaging in political campaigning or lobbying, including generating campaign materials personalized to or targeted at specific demographics"); Anthropic AUP ("Do Not Create Political Campaigns or Interfere in Elections").

[73] OpenAI AUP ("Engaging in or promoting academic dishonesty"); Anthropic AUP ("Plagiarize or engage in academic dishonesty").

[74] OpenAI AUP ("Don't build tools that target users under 13 years of age"); Anthropic AUP ("Do Not Compromise Children's Safety"); Meta AUP ("The illegal distribution of information or materials to minors, including obscene materials, or failure to employ legally required age-gating in connection with such information or materials"); Google AUP ("Promoting or generating content related to child sexual abuse or exploitation").

[75] OpenAI AUP ("Providing tailored legal, medical/health, or financial advice without review by a qualified professional and disclosure of the use of AI assistance and its potential limitations"); Anthropic AUP ("Human-in-the-loop: when using our products or services to provide advice, recommendations, or subjective decisions that directly impact individuals in high-risk domains, a qualified professional in that field must review the content or decision prior to dissemination or finalization"); Meta AUP ("Engage in the unauthorized or unlicensed practice of any profession including, but not limited to, financial, legal, medical/health, or related professional practices"); Google AUP ("Engaging in the unauthorized or unlicensed practice of any profession including, but not limited to, financial, legal, medical/health, or related professional practices").

[76] *See, e.g.*, Llama 3.1 Community License Agreement *supra* note 52 (requiring model developers building on top of its open weight Llama model to "include 'Llama 3' at the beginning of any such AI model name.").

Other closed-access model providers have used the terms to intimidate researchers investigating model vulnerabilities.[79] In one case, Midjourney (a closed model provider specializing in image generation models) had Terms of Service that bluntly stated, "If You knowingly infringe someone else's intellectual property, and that costs us money, we're going to come find You and collect that money from You. We might also do other stuff, like try to get a court to make You pay our attorney's fees. Don't do it."[80] This came at a time when researchers investigated whether models could generate copyright infringing material. The company removed this term after researchers raised concerns about the chilling effects such a term has.[81]

Blocking access to the account works if an account is needed to access the model. But when model weights are released openly, the only way for a model creator to enforce these terms is by monitoring for misuse and then seeking legal recourse. Meta has emphasized the important role that these terms play in its strategy for responsibly releasing model weights. In 2024, for example, when the National Telecommunications and Information Administration ("NTIA") sought request for responses to the question, "Could open foundation models reduce equity in rights and safety-impacting

---

[77] Elizabeth Dwoskin, *OpenAI suspends bot developer for presidential hopeful Dean Phillips*, WASH. POST (Jan. 20, 2024), https://www.washingtonpost.com/technology/2024/01/20/openai-dean-phillips-ban-chatgpt/.

[78] Benj Edwards, *Ban Warnings Fly as Users Dare to Probe the "Thoughts" of OpenAI's Latest Model*, ARS TECHNICA (Sept. 16, 2024, 6:49 PM), https://arstechnica.com/information-technology/2024/09/openai-threatens-bans-for-probing-new-ai-models-reasoning-process/.

[79] Shayne Longpre et al., Position: A Safe Harbor for AI Evaluation and Red Teaming, in Proceedings of the Forty-First International Conference on Machine Learning (2023).

[80] Terms of Service, Midjourney (December 22, 2023), https://web.archive.org/web/20240226144923/https://docs.midjourney.com/docs/terms-of-service

[81] Nitasha Tiku, *Top AI researchers say OpenAI, Meta and more hinder independent evaluations*, WASH. POST (Mar. 5, 2024), https://www.washingtonpost.com/technology/2024/03/05/ai-research-letter-openai-meta-midjourney/

AI systems (e.g. healthcare, education, criminal justice, housing, online platforms, etc.)?"[82] Meta (an open weight model provider) responded that, "We developed and made available responsible use guides to help downstream developers measure impact. And it's why the Llama 2 Acceptable Use Policy prohibits specific high risk use cases."[83] In response to the same question, Stability AI (which provides open weight image generation models) stated:

> As a first line of defense, models may be optimized for safer behavior prior to release through a range of techniques including data curation, instruction tuning, reinforcement learning from human or AI feedback, or direct policy optimization. For example, Stability AI filters unsafe content from training data, helping to prevent the model from producing unsafe content. Following pre-training, we evaluate and fine-tune our models to help eliminate undesirable behaviors, such as unacceptable bias. We disclose known risks and limitations in standardized formats, such as model cards, to help downstream deployers decide on additional mitigations. Our most capable models are subject to acceptable use licenses that prohibit a range of unlawful or misleading applications.[84]

Open model providers have made some attempts to get compliance with its terms. Meta initially issued DMCA takedown requests to unofficial repositories hosting its Llama model,[85] though eventually they gave up as the model proliferated online and parties began contesting the takedown requests.[86] To our knowledge no model creator has, as of yet, sought to legally enforce these responsible use terms, even in the face of clear evidence of downstream misuse. But given the actual and potential misuse of AI models by users and the use by competitors training their own models, efforts to enforce these terms of service in court are likely on their way. In the next

---

[82] Dual Use Foundation Artificial Intelligence Models With Widely Available Model Weights, 89 Fed. Reg. 14059, 14062 (Feb. 26, 2024)

[83] Meta Platforms, Inc., Comment Letter on Dual Use Foundation Artificial Intelligence Models With Widely Available Model Weights (Mar. 2024), https://about.fb.com/wp-content/uploads/2024/03/NTIA-RFC-Meta-Response-March-2024.pdf

[84] Stability AI Comment Letter on Dual Use Foundation Artificial Intelligence Models With Widely Available Model Weights (Mar. 2024), https://www.regulations.gov/comment/NTIA-2023-0009-0333

[85] Notice of Claimed Infringement via Email, Github (Mar. 20, 2023), https://github.com/github/dmca/blob/280652a060b86de87f223737ae54307a292fb96b/2023/03/2023-03-21-meta.md

[86] *Id.*

parts, we discuss how those efforts might fare.

## II. Copyright in Underlying Assets

In most contexts, conditional restrictions on the use of copyrighted material are enforceable through copyright law. For instance, a software developer might release code under a license limiting its use to non-commercial purposes. Should a competitor violate these terms, the developer could pursue copyright infringement claims against acts that would violate the copyright but for the license.[87] This is not a novel legal concept; many developers enforce their software licenses through copyright infringement litigation.

Generative AI models, though, are not like traditional software. Despite their value, it is not clear that the user-accessible parts of those systems (outputs and model weights) are copyrightable at all. Recent Copyright Office guidance and case law suggest that generative AI model creators are unlikely to be considered authors of model outputs.[88] And most disclaim ownership of them.[89] Even would-be users of models, despite their efforts to the contrary, face a high bar to be considered owners of AI model outputs.[90]

Model weights likely fare no better. They are also created through an automated process, likely rendering them uncopyrightable, like their outputs.[91] And even if the human authorship requirement is met, model weights are likely functional artifacts unprotected by copyright.[92]

Model creators might claim ownership over some of the training data, and then claim that models are a copyrightable database of that data—but this too

---

[87] See, e.g., Jacobsen v. Katzer, 535 F.3d 1373, 1383 (Fed. Cir. 2008) (holding that terms in open-source software licenses are "enforceable copyright conditions"); MDY Industries, LLC v. Blizzard Entertainment, Inc., 629 F.3d 928, 941 (9th Cir. 2010) (ruling that violating Terms of Use attached to a software license can constitute copyright infringement if there is a nexus between the condition and the licensor's copyright rights).

Though, the legal rule is more complex than the text suggests. The law distinguishes between "covenants" and "conditions" in copyright licenses. Only limits that are tied directly to the grant of the copyright can be enforced by a copyright suit. Other restrictions in a copyright license, called covenants, may be enforceable as a matter of contract law but not in a suit for copyright infringement. *See, e.g.*, *Sun Microsystems v. Microsoft Corp.*, 188 F.3d 1115, 1116-18 (9th Cir. 1999). We discuss contract enforcement of this sort in Part III.

[88] Copyright Registration Guidance: Works Containing Material Generated by Artificial Intelligence, 88 Fed. Reg. 16190 (Mar. 16, 2023) [hereafter "Copyright Office AI Authorship Guidance"]; see also Thaler v. Perlmutter, No. 22-cv-01564 (BAH), 2023 WL 5333435 (D.D.C. Aug. 18, 2023).

[89] *See* Part II.A.

[90] *See supra* note 88.

[91] *See* Part II.B.

[92] *See* Part II.C.

is tenuous. It is true that model weights might occasionally store memorized versions of copyrighted training data, which might be protectable.[93] But model creators work diligently to prevent this.[94] And they tend to argue in court that this is atypical or impossible.[95]

Consequently, generative AI companies may lack key intellectual property rights in their product, limiting their ability to enforce terms on downstream model users through copyright infringement litigation. This Part examines the underlying issues preventing such claims, considering potential copyright infringement arguments related to: (1) model outputs; (2) model weights; (3) encoded copies or derivative works of copyrighted training data; and (4) system code required for model operation. Each of these arguments faces significant legal hurdles.

### A. Model Outputs

To the extent that model creators think that there is any copyright ownership to be had over model outputs, they appear to assign those rights to their users—conditional on compliance with the terms of use. Table 2 provides an overview of these terms for various model providers.

*Table 2. Terms related to ownership of content.*

| OpenAI Consumer Terms | **Ownership of content.** As between you and OpenAI, and to the extent permitted by applicable law, you (a) retain your ownership rights in Input and (b) own the Output. We hereby assign to you all our right, title, and interest, if any, in and to |
| --- | --- |

---

[93] *See, e.g.*, Nicholas Carlini et al., Extracting Training Data from Large Language Models, 2021 USENIX Security Symposium 2633 (2021) (demonstrating that large language models can unintentionally memorize and reproduce training data, including sensitive personal information).

[94] For examples of techniques for preventing infringement in model outputs *see* Henderson et al., Foundation Models and Fair Use, 24 J. Machine Learning Rsch. 1 (2023); Edward Lee et al., Talkin' 'Bout AI Generation: Copyright and the Generative-AI Supply Chain, 71 J. Copyright Soc'y U.S.A. 1 (2023); Matthew Sag, Copyright Safety for Generative AI, 61 Hous. L. Rev. 295 (2023).

[95] See, e.g., Memorandum of Law in Support of OpenAI Defendants' Motion to Dismiss at 11, N.Y. Times Co. v. Microsoft Corp., No. 1:23-cv-11195 (S.D.N.Y. filed Feb. 26, 2024) (arguing that the alleged reproduction of copyrighted content was "highly anomalous" and resulted from "targeting and exploiting a bug" through "deceptive prompts that blatantly violate OpenAI's terms of use," and asserting that "[n]ormal people do not use OpenAI's products in this way").

| | |
|---|---|
| | Output.<br><br>**Similarity of content.** Due to the nature of our Services and artificial intelligence generally, output may not be unique and other users may receive similar output from our Services. Our assignment above does not extend to other users' output or any Third Party Output.<br><br>**Our use of content.** We may use Content to provide, maintain, develop, and improve our Services, comply with applicable law, enforce our terms and policies, and keep our Services safe.[96] |
| Anthropic | "As between you and Anthropic, and to the extent permitted by applicable law, you retain any right, title, and interest that you have in the Prompts you submit. Subject to your compliance with our Terms, we assign to you all of our right, title, and interest—if any—in Outputs."[97] |
| Meta Llama 3 | "Subject to Meta's ownership of Llama Materials and derivatives made by or for Meta, with respect to any derivative works and modifications of the Llama Materials that are made by you, as between you and Meta, you are and will be the owner of such derivative works and modifications."[98] |
| Google Gemini | "Some of our Services allow you to generate original content. Google |

---

[96] OpenAI Terms of Use *supra* note 50.
[97] Anthropic Consumer Terms of Service *supra* note 51.
[98] Meta Llama 3 Community License Agreement, Meta (Apr. 18, 2024), https://www.llama.com/llama3/license/

| | won't claim ownership over that content. You acknowledge that Google may generate the same or similar content for others and that we reserve all rights to do so."[99] |
|---|---|
| Google Gemma (open weight model) | "Google claims no rights in Outputs you generate using Gemma. You and your users are solely responsible for Outputs and their subsequent uses."[100] |
| Midjourney (closed weight model) | You own all Assets You create with the Services to the fullest extent possible under applicable law. There are some exceptions:<br>• Your ownership is subject to any obligations imposed by this Agreement and the rights of any third-parties.<br>• If you are a company or any employee of a company with more than $1,000,000 USD a year in revenue, you must be subscribed to a "Pro" or "Mega" plan to own Your Assets.<br>• If you upscale the images of others, these images remain owned by the original creators.[101] |
| Stable Diffusion 3 Medium | As between You and Stability AI, You own any outputs generated from the Models or Derivative Works to the extent permitted by applicable law. [102] |

This conditional transfer of rights sometimes contrasts with external statements by companies. For example, one representative of OpenAI suggested that the transfer of ownership is unconditional: "OpenAI will not claim copyright over content generated by the API for you or your end users."[103]

---

[99] Gemini API Additional Terms of Service, Google (May 14, 2024), https://ai.google.dev/gemini-api/terms
[100] Gemma Terms of Use, Google (Apr. 1, 2024), https://ai.google.dev/gemma/terms
[101] Midjourney Terms of Service *supra* note 80.
[102] License for Stable Diffusion 3 Medium Model *supra* note 49

This assignment of ownership is somewhat inconsequential; current law is unlikely to recognize model creators as authors of generated outputs.[104] In March 2023, the Copyright Office issued a statement of policy clarifying its practices for examining and registering works that contain material generated by artificial intelligence.[105] The statement provides guidance regarding its application of copyright law's human authorship requirement to such works. It cited Supreme Court case law defining an "author" as a human being[106] and lower court decisions holding that non-human expression is ineligible for copyright.[107] Accordingly, the Office stated that it will not register works "produced by a machine or mere mechanical process that operates randomly or automatically without any creative input or intervention from a human author."[108]

However, human-authored works containing AI-generated components could be registered as long as the human-authored aspects "contain a sufficient amount of original, creative authorship."[109] For example, a journalist who writes an article that includes some AI-generated text can claim copyright in their article as a whole and the human-written portions. But the AI-generated portions must be disclaimed, and applicants have a duty to disclose its inclusion in the application.[110] If anyone is a co-author and owner of the model outputs, therefore, it is likely to be the users, not the model creators.[111]

The Office also explicitly suggested that "prompting" a model (providing instructions on how to generate a piece of content), will often not be registrable.[112] For example, when an AI technology receives solely a prompt from a human and produces complex written, visual, or musical works in response, the Office explained that "the 'traditional elements of authorship' are determined and executed by the technology—not the human user."[113] This is because, based on the Office's current understanding of generative AI technologies, "users do not exercise ultimate creative control over how such systems interpret prompts and generate material."[114]

---

[103] *Will OpenAI Claim Copyright Over What Outputs I Generate with the API?*, OpenAI Help Ctr., https://help.openai.com/en/articles/5008634-will-openai-claim-copyright-over-what-outputs-i-generate-with-the-api (last visited Oct. 24, 2024).

[104] *See also* Mark A. Lemley, *How Generative AI Turns Copyright Law Upside Down*, 25 SCI. & TECH. L. REV. 190 (2024).

[105] Copyright Office AI Authorship Guidance *supra* note 88.

[106] *Id.* at 16191 (citing Burrow-Giles Lithographic Co. v. Sarony, 111 U.S. 53, 58 (1884)).

[107] *Id.* (citing Urantia Found. v. Kristen Maaherra, 114 F.3d 955, 957-59 (9th Cir. 1997) and Naruto v. Slater, 888 F.3d 418, 426 (9th Cir. 2018)).

[108] *Id.* (quoting U.S. Copyright Office, *Compendium of U.S. Copyright Office Practices* sec. 313.2 (3d ed. 2021) (" *Compendium (Third)"*)).

[109] *Id.*

[110] *Id.*

[111] *See also* Lemley *supra* note 104.

[112] Copyright Office AI Authorship Guidance *supra* note 88 at 16192.

The Office analogized these prompts to "instructions to a commissioned artist" in that "they identify what the prompter wishes to have depicted, but the machine determines how those instructions are implemented in its output."[115] It provided the example of a user instructing an AI system to "write a poem about copyright law in the style of William Shakespeare." While the user can expect the resulting output to "be recognizable as a poem, mention[] copyright, and resemble[] Shakespeare's style," it is the "technology [that] will decide the rhyming pattern, the words in each line, and the structure of the text."[116] In such a case, the Office concluded that the "generated material is not the product of human authorship" and thus not protectable by copyright.[117] Ultimately, for a generative AI work to qualify for protection, the Office stated that a human must have exercised "creative control" over the work and "actually formed" its traditional elements of authorship.[118]

This appears to leave little room for authors leveraging prompt-based approaches to yield anything but a compilation right in a work—and even then, only if they heavily modify the outputs *ex post* to qualify as a compilation.[119] The Copyright Office has stuck to this policy position, rejecting a number of requests for registration for prompt-based generated material,[120] except for at least one recent case which was given a compilation

---

[113] *Id.*
[114] *Id.*
[115] *Id.*
[116] *Id.*
[117] *Id.*
[118] *Id.*
[119] Lemley, *Upside Down*, *supra* note 104.
[120] See U.S. Copyright Office Review Board, Re: Reconsideration of Second Request for Reconsideration for Refusal to Register A Recent Entrance to Paradise, SR # 1-7100387071 (Feb. 14, 2022), https://www.copyright.gov/rulings-filings/review-board/docs/a-recent-entrance-to-paradise.pdf (refusing to register a work created autonomously by artificial intelligence without any creative contribution from a human actor); see also U.S. Copyright Office Review Board, Re: Registration Decision Regarding "Zarya of the Dawn," Correspondence ID: 1-3ZPC6C3 (Feb. 21, 2023), https://www.copyright.gov/docs/zarya-of-the-dawn.pdf (concluding that individual AI-generated images in a graphic novel were not protectable under copyright law, while allowing registration for the overall work as a compilation); U.S. Copyright Office Review

copyright based on heavy *ex post* modifications by the human author.[121]

As one of us has noted, in other cases it may be that prompts, if sufficiently creative and the core of any expressive copyrightable work, are the copyrightable and protectable aspect of the prompt-to-output pipeline.[122] But if so, it is the user's contribution, not the AI's, that qualifies it for copyright protection. At most, AI companies might get a license to the *user's* copyright, of which the outputs are a derivative work. And any work generated purely as a function of their AI model is not copyrightable.[123]

Model creators might be selling model outputs, but they don't own them and cannot claim infringement for users copying those outputs.

### B. Model Weights

Model creators' claims of ownership over model weights would fare no better. Recall that model weights, in essence, are series of numerical values that, when applied to input data, produce the model's output. These weights are typically stored in files that contain a series of numbers. In layman's terms might read something like, "Matrix Number 5 in Layer 3 of the overall model architecture should contain values: [0.1, 1.2, …]." These weights are "learned" through a semi-automated process involving "training" the model on vast amounts of data.[124] While model creators can exert significant control over downstream model behaviors, at the scale of billions of parameters

---

Board, Re: Registration Decision Regarding SURYAST, Correspondence ID: 1-5DM8ZFK (Dec. 11, 2023), https://copyright.gov/rulings-filings/review-board/docs/SURYAST.pdf; U.S. Copyright Office Review Board, Re: Registration Decision Regarding Théâtre D'opéra Spatial, Correspondence ID: 1-5JJHC8M (Sep. 5, 2023), https://www.copyright.gov/rulings-filings/review-board/docs/Theatre-Dopera-Spatial.pdf.

[121] See U.S. Copyright Office Review Board, Re: Registration Decision Regarding "Zarya of the Dawn," Correspondence ID: 1-3ZPC6C3 (Feb. 21, 2023), https://www.copyright.gov/docs/zarya-of-the-dawn.pdf (allowing registration of a graphic novel containing AI-generated images as a compilation authorship while denying protection for the individual AI images themselves).

[122] Lemley, *supra* note 104.

[123] The United States Patent and Trademark Office ("USPTO") takes a slightly different position for AI inventorship and has explicitly stated that humans who develop an "essential building block" from which the AI derives the invention may qualify as inventors—potentially including the "natural person(s) who design[], build[], or train[] an AI system in view of a specific problem to elicit a particular solution." But one must go beyond simply "owning or overseeing and AI system" or having "intellectual domination" over the AI system to claim inventorship in the creation—potentially excluding general purpose model creators. USPTO, Inventorship Guidance for AI-Assisted Inventions, 89 Fed. Reg. 10043 (Feb. 13, 2024).

[124] See Ian Goodfellow et al., Deep Learning 267-309 (2016) (describing the process of training neural networks and determining optimal weights).

(collectively, weights), model creators do not exert fine-grained control over the individual numbers in the set. They are not written by human authors in the way that, say, computer code is.

We think that, like model outputs, model weights are likely to be unprotectable in most cases.[125] The copyrightability concerns surrounding model weights bear similarities to those of generative AI outputs. Just as the Copyright Office determined that much purely AI-generated content lacks the necessary human authorship for copyright protection,[126] the same reasoning may apply to model weights. The process of creating model weights, like generating AI outputs, is highly stochastic and semi-automated, as we discuss below.

The Copyright Office has stated that for a work to be copyrightable, a human must have exercised "creative control" over the work and "actually formed" its traditional elements of authorship.[127] In the case of model weights, as with AI-generated outputs, the final result is largely determined by algorithmic processes rather than direct human creativity.[128]

The Office's analogy of AI prompts to "instructions to a commissioned artist" could equally apply to the process of training model weights, where human input is limited to only some high-level interventions, like modifying the training data and methods. Humans do have *some* level of control over model outputs—researchers have trained models that are designed to take on a particular viewpoint about political issues, for example.[129] To do so, model

---

[125] *See also* Sancho McCann, *Copyright Throughout a Creative AI Pipeline*, 19 AN. J.L. & TECH. 109, 120 (2023) (noting similar challenges around the protectability of model weights in the Canadian legal context); Begoña Gonzalez Otero, *Machine Learning Models Under the Copyright Microscope: Is EU Copyright Fit for Purpose?,* 70 GRUR J. OF EUROPEAN AND INTERN'L IP LAW 1043 (2021) (arguing similarly in the EU context); Hao-Yun Chen, Copyright Protection for Software 2.0? Rethinking the Justification of Software Protection under Copyright Law, in Artificial Intelligence and Intellectual Property 283 (2021); Peter R. Slowinski, Rethinking Software Protection, in Artificial Intelligence & Intellectual Property 341 (2021).

[126] U.S. Copyright Office, Copyright Registration Guidance: Works Containing Material Generated by Artificial Intelligence (Mar. 16, 2023), https://www.copyright.gov/ai/ai_policy_guidance.pdf

[127] *Id.*

[128] Though, in some cases, model creators *can* introduce fine-grained edits to very particular model behaviors, this only targets a subset of model behaviors. The majority of behaviors are learned in an automated fashion. *See, e.g.*, Adly Templeton et al., Scaling Monosemanticity: Extracting Interpretable Features from Claude 3 Sonnet, Transformer Circuits (May 21, 2024), https://transformer-circuits.pub/2024/scaling-monosemanticity/index.html (describing a method for fine-grained control over certain interpretable model behaviors); Eric Mitchell et al., Fast Model Editing at Scale, in Proc. of the Tenth Int'l Conf. on Learning Representations (2022).

[129] *See, e.g.*, Dominik Stammbach, Philine Widmer, Eunjung Cho, Caglar Gulcehre &

creators typically gather a customized dataset reflecting the desired perspectives and update the model weights until the model is more likely to generate outputs consistent with those views. The level of human intervention is equally, or perhaps more, abstract than in the case of AI prompts. The specific values of the weights are determined autonomously by the training process, guided to some extent by the selected data. This lack of direct human authorship in the creation of model weights strongly suggests that they, like AI-generated outputs, may fall outside the scope of copyright protection under the status quo.[130]

The copyrightability of model weights is further complicated by their functional nature. 17 U.S.C. § 102(b) states that copyright protections do not extend to "any idea, procedure, process, system, method of operation, concept, principle, or discovery, regardless of the form in which it is described, explained, illustrated, or embodied in such work." Courts have consistently held that purely functional aspects of software are not protected by copyright.[131]

Generally, in these cases defendants were copying some functional element of the original program, rather than the program as a whole. For example, they might copy just the interface of a program to ensure interoperability, while completely replacing the expressive guts of the system. By contrast, downstream users would presumably download and copy the model weights verbatim. Some might argue that that makes a difference. Literal copying of object code (the 1s and 0s in which a program is implemented in a computer) is copyright infringement even though the code isn't human-readable.[132]

---

Elliott Ash, Aligning Large Language Models with Diverse Political Viewpoints (June 2024) (unpublished manuscript) (available at https://arxiv.org/abs/2406.14155).

[130] Some argue that this limited guidance of model outputs creates implications for First Amendment protections. *See, e.g.*, Mackenzie Austin & Max Levy, *Speech Certainty: Algorithmic Speech and the Limits of the First Amendment*, 77 STAN. L. REV. ___ (forthcoming 2025); Peter Salib, *AI Outputs Are Not Protected Speech*, 97 WASH. U. L. REV. ___ (forthcoming 2024). We do not address the First Amendment question here, but note that the level of human control over model outputs to receive constitutional protections may be different than the level of control needed to receive copyright ownership in the model and its outputs.

[131] See Computer Assocs. Int'l, Inc. v. Altai, Inc., 982 F.2d 693, 703-04 (2d Cir. 1992) (holding "those elements of a computer program that are necessarily incidental to its function are" unprotectible); SAS Institute Inc. v. World Programming Ltd., 64 F. Supp. 3d 755, 776 (E.D.N.C. 2014) (finding that a programming language is an idea that is not copyrightable) (affirmed 952 F.3d 513 (4th Cir. 2020)); Pamela Samuelson, *Functionality and Expression in Computer Programs: Refining the Tests for Software Copyright Infringement*, 31 BERKELEY TECH. L.J. 1215, 1220-21 (2016).

[132] Apple Computer v. Franklin Computer Corp., 714 F.2d 1240, 1243 (3d Cir. 1983)

Nonetheless, we think copying model weights is not like copying the entirety of a computer program in object code. Object code is (for the most part) derived from a human-written copyrightable expression of an idea. Programmers hand-craft code in high-level programming languages, which is then converted into object code through a "compiler." Model weights are not programmed by humans in the same way and are not a clear-cut reflection of human authorship, like object code is.

## C. Training Data

Finally, one could argue that model weights are a protectable database or a compilation encoding their training dataset.[133] But this is an uphill battle. AI companies generally don't own the underlying data that they train on. Rather, they scrape it from a variety of sources, often publicly available internet data.[134] That process is itself the subject of dozens of legal challenges from the copyright owners over their use of data.[135]

Perhaps the aggregation of that data could be copyrighted even though

---

("There are three levels of computer language in which computer programs may be written. High level language, such as the commonly used BASIC or FORTRAN, uses English words and symbols, and is relatively easy to learn and understand (e.g., 'GO TO 40' tells the computer to skip intervening steps and go to the step at line 40). A somewhat lower level language is assembly language, which consists of alphanumeric labels (e.g., 'ADC' means 'add with carry'). Statements in high level language, and apparently also statements in assembly language, are referred to as written in 'source code.' The third, or lowest level computer language, is machine language, a binary language using two symbols, 0 and 1, to indicate an open or closed switch (e.g., '01101001' means, to the Apple, add two numbers and save the result). Statements in machine language are referred to as written in 'object code.' The CPU can only follow instructions written in object code. However, programs are usually written in source code which is more intelligible to humans. Programs written in source code can be converted or translated by a 'compiler' program into object code for use by the computer. Programs are generally distributed only in their object code version stored on a memory device. ") (cleaned up).

[133] There is some existing litigation around scraping of copyrighted "automated databases." *See, e.g,*. Complaint, Jobiak, LLC v. Botmakers LLC, No. 2:23-cv-8604 (C.D. Cal. filed Oct. 12, 2023)

[134] *See, e.g.*, Shayne Longpre et al., The Responsible Foundation Model Development Cheatsheet: A Review of Tools & Resources (Mar. 26, 2024) (unpublished manuscript), https://arxiv.org/abs/2406.16746 (surveying popular large-scale datasets, including various webcrawls, journal articles, and more); Yang Liu et al., Datasets for Large Language Models: A Comprehensive Survey (Feb. 28, 2024) (unpublished manuscript), https://arxiv.org/abs/2402.18041 (conducting a similar survey); Colin Raffel et al., Exploring the Limits of Transfer Learning with a Unified Text-to-Text Transformer (Oct. 23, 2019) (unpublished manuscript), https://arxiv.org/abs/1910.10683 (introducing "C4" a popular dataset of websites based on another corpus named "CommonCrawl").

[135] *See, e.g.*, Complaint, Authors Guild v. OpenAI, Inc., No. 1:23-cv-08292 (S.D.N.Y. filed Sept. 19, 2023) (alleging copyright infringement in AI training data)

the AI company doesn't own any of the individual data itself? We are skeptical. The United States has no freestanding database right, and while copyright has been granted in the arrangement and compilation of works there must be sufficient creativity in development of that compilation.[136] The copyright also does not extend independently to the data within the database.[137] It is not clear that the dataset itself has enough creativity and originality associated with it to be protectable.

Most model creators use a series of automated mechanisms to scrape, process, and filter data to craft the original datasets. And most datasets are themselves derivatives of other datasets. For example, many model creators start with Common Crawl and then filter the data for quality.[138] The selection of "everything available for scraping on the internet" is not sufficiently creative to qualify for copyright protection. If there is enough creativity for an independent compilation right, it is murky at best, and would be limited to the protection of deliberate individualized filtering decisions.

Even if datasets merit copyright protection as compilations due to their curatorial decisions, the relationship between training data and model outputs means that the output is not likely to be a derivative work of the dataset. Machine learning models do not necessarily "learn" from datasets holistically in the way compilation rights contemplate. Rather, the influence of any given training example may vary dramatically based on model architecture, training procedure, and interactions with other data points.

Determining when and whether protectable elements of individual works are meaningfully encoded in a model remains unsolved—except for clearcut cases like when a model memorizes a particular training point. Due to the vast scale of training data and phenomena like catastrophic forgetting,[139] individual examples may significantly shape model functionality or have negligible impact. This renders traditional tests for derivative works poorly suited to AI models trained on large datasets. To be a derivative work, the defendant's product (here, the model itself) must be "substantially similar" to a particular underlying original.[140] That will only rarely be true,[141] and if it is, it is more straightforward to make a claim based on copying.

---

[136] *See, e.g.,* Feist Publications, Inc. v. Rural Telephone Service Co., 499 U.S. 340 (1991) (holding that factual compilations like telephone directories lack sufficient originality for copyright protection, establishing the "modicum of creativity" standard); Assessment Technologies of WI, LLC v. WIREdata, Inc., 350 F.3d 640 (7th Cir. 2003) (ruling that a database with enough originality can be copyrightable, but that copyright does not extend to the public domain data contained within it).

[137] *Feist*, 499 U.S. at 340.

[138] *See, e.g.,* Stefan Baack, *A Critical Analysis of the Largest Source for Generative AI Training Data: Common Crawl*, in Proc. 2024 ACM Conf. on Fairness, Accountability & Transparency 2199 (2024) (describing how Common Crawl is used as part of a variety of different Generative AI datasets).

[139] Where the model completely forgets anything about a datapoint after seeing too many other datapoints. *See, e.g.*, James Kirkpatrick et al., *Overcoming Catastrophic Forgetting in Neural Networks*, 114 PROC. NAT'L ACAD. SCI. 3521, 3521-22 (2017) (describing how neural networks can completely lose previously learned information when trained on new tasks)

A final possibility involves AI companies that create their own data for training. Models have been shown to memorize some datapoints completely in rare cases.[142] Model creators could argue that the model is not a database, but rather a compressed format for storing some of their works.[143] A model creator could create data that they own the copyright to force the model to memorize this content, and then argue that the model is just a way to access this data. Since they own the copyright to the memorized data, they can attach terms to it. It seems unlikely that model creators would make this argument in court. Not only is it technically dubious, but it is also the very argument copyright owners are making *against* AI companies to allege that the models themselves are illegal. It would be against the model creators' interests to try to establish that the models are all derivative works of the content on which they are trained.[144]

---

[140] *See, e.g.*, Litchfield v. Spielberg, 736 F.2d 1352, 1357 (9th Cir. 1984) ("We have stated that a work will be considered a derivative work only if it would be considered an infringing work if the material which it has derived from a prior work had been taken without the consent of a copyright proprietor of such prior work. To prove infringement, one must show substantial similarity.") (cleaned up); Andersen v. Stability AI Ltd., 23-cv-00201-WHO, 12 (N.D. Cal. Oct. 30, 2023) (using a similar rule in an AI context: "I am not convinced that copyright claims based a derivative theory can survive absent "substantial similarity" type allegations").

[141] *See, e.g.*, Albert Ziegler, *GitHub Copilot: Parrot or Crow? A First Look at Rote Learning in GitHub Copilot Suggestions*, GITHUB BLOG (June 30, 2021), https://github.blog/ai-and-ml/github-copilot/github-copilot-research-recitation/ (finding that in the early days of Github's CoPilot about one "recitation event" happened every 10 user weeks); Percy Liang et al., Holistic Evaluation of Language Models, in Transactions on Machine Learning Rsch. (2023) (similarly finding low rates of regurgitation except for very specific pieces of content); Peter Henderson et al., *Foundation Models and Fair Use*, 24 J. MACHINE LEARNING RSCH. 400 (2023) (finding similarly low rates of regurgitation, except in specific cases of some memorized content).

[142] *See* Nicholas Carlini et al., Extracting Training Data from Large Language Models, 2021 USENIX Security Symposium 2633, 2634 (2021) (demonstrating that large language models can unintentionally memorize and reproduce training data)

[143] *See* A Feder Cooper & James Grimmelmann, *The Files Are in the Computer: Copyright, Memorization, and Generative AI* (Apr. 19, 2024) (unpublished manuscript), https://arxiv.org/abs/2404.12590 (arguing that when a model memorizes a piece of content it is stored "in" the model).

[144] For a court rejection of that derivative works claim, see Andersen v. Stability AI Ltd., 23-cv-00201-WHO, 13 (N.D. Cal. Oct. 30, 2023) ("Defendants make a strong case that I should dismiss the derivative work theory without leave to amend because plaintiffs cannot plausibly allege the Output Images are substantially similar or re-present protected aspects of copyrighted Training Images, especially in light of plaintiffs' admission that Output

### D. System Code

The final, and most likely, source of copyright protection is the software that runs the model itself. Models are typically released with a thin layer of software that allows users to execute the process encoded by model weights. Some companies may even release the code used to train the model. We will refer to this entire set of code as "system code." We will refer to the code used to train a model as "training system code" and the code used to execute instructions encoded by the model weights as "inference system code."[145] It is possible (though not guaranteed) that both of these codebases may be protectable by copyright. The training system code is generally not implicated for the terms of service attached to models, so we will not account for it here.

Model creators may argue that inference system code is copyrightable, and that by using that code to read and run inference on model weights, downstream users are liable for infringement if they do not comply with associated terms. This is a plausible argument, but for two practical problems.

First, in their current form, the software required to execute model weights is relatively simple and purposefully interoperable. Model users can simply switch to a different software to decode model weights. Many codebases can load and execute released model weights, using an interchangeable application program interface ("API").

Second, even if a model creator purposefully made model weights purposefully non-interoperable, downstream users would find a way to reverse engineer this process to create compatible execution software. Doing so is exactly what the Court in *Google v. Oracle* ruled was fair use.[146] Recently, the court in *SAS Institute Inc. v. World Programming Ltd.*,[147] would go even further and argue that these application program interfaces are unprotectable, not even reaching the fair use argument. And merger doctrine could also exclude copyrightability in some jurisdictions.[148]

---

Images are unlikely to look like the Training Images. . . But other parts of plaintiffs' Complaint allege that Output Images can be so similar to plaintiff's styles or artistic identities to be misconstrued as 'fakes' . . . Once plaintiffs amend, hopefully providing clarified theories and plausible facts, this argument may be re-raised on a subsequent motion to dismiss.") (cleaned up).

[145] In machine learning parlance, "training" creates the model weights, while "inference" is the action of executing the instructions encoded by the model weights.

[146] Google LLC v. Oracle Am., Inc., 141 S. Ct. 1183, 1190 (2021).

[147] 64 F. Supp. 3d 755, 776 (E.D.N.C. 2014) (affirmed 64 F. Supp. 3d 755 (E.D.N.C. 2014)).

[148] Lexmark Intern. v. Static Control Components, 387 F.3d 522, 536 (6th Cir. 2004)

Overall, the implementing code is not the important part of the AI model, so while it is likely copyrightable, competitors generally won't need to copy that code to make use of the model. And if they do, they will be free to reverse-engineer it.[149]

### III. Contract Claims

Without a viable copyright claim, model creators will no doubt look to contract claims. Unfortunately for AI companies, these too are not so straightforward. Contract claims based on failed copyright arguments may fall to preemption, and there are limits on the usefulness of contract law even if it does apply. We note that even if terms of service are likely to be enforceable for some narrow "responsible use" terms, anti-competitive provisions and many restraints on copying will likely be thrown out.

#### A. *Are Terms of Use Enforceable Contracts at All?*

The first question is whether AI company terms of use create an enforceable contract at all.

AI companies invariably take the position that "you need to come to us to access it, so we can impose whatever restrictions we want." That is generally true as to access itself—even if OpenAI does not require a login, they can refuse to allow anyone to use ChatGPT if they want. But assuming they have not blocked a particular user, is a user bound to an AI company's terms and conditions merely by reason of having used the site?

Most U.S. courts answer this question in the affirmative.[150] The Restatement of Consumer Contracts takes the rather extreme position that visiting a site itself means accepting whatever terms the site owner establishes.[151] The case law has not gone that far. Most courts require that the consumer be put on reasonable notice of the existence of the terms on the front page of the site, rather than having to search or scroll down to find the link to the terms, and that the link to the terms be conspicuous.[152] That is not a very serious requirement, and one of us has argued elsewhere that it doesn't reflect a contract in any traditional sense and permits a great deal of

---

("to the extent compatibility requires that a particular code sequence be included in the component device to permit its use, the merger and scènes à faire doctrines generally preclude the code sequence from obtaining copyright protection.")

[149] The act of reverse-engineering the code may lead to other claims aside from infringement that we will discuss later in this work. *See* Part III.B.

[150] *See, e.g.,* Toth v. Everly Well, Inc., __ F.4th __ (1st Cir. Sept. 25, 2024); Mark A. Lemley, *Terms of Use*, (collecting cases).

[151] Restatement of Consumer Contracts § 2 at 30.

[152] Lemley *supra* note 27

mischief.[153]

Many AI companies have begun to meet this low barrier. ChatGPT contains a blue hyperlink on the home screen to its terms and conditions (and—admirably—writes those terms in an understandable way).[154] Anthropic's Claude currently requires accepting terms to create an account.[155] And to download Meta's open-weight Llama models or Stability AI's stable-diffusion models from their official repositories, one now has to agree to a clickwrap agreement.

However, many open-weight models are reshared in ways that do not require agreeing to anything at all. Stable Diffusion's models, for example, can be accessed through a third-party interface that does not have terms anywhere on the page.[156] And models are regularly re-uploaded with license terms modified or omitted.

But even for companies that have followed the minimal procedures to make their terms of service enforceable under current law, there may be limits on their right to condition provision of the service on improper things. It is questionable whether they could say "no competitors allowed"; we discuss that issue below.[157] More generally, contract law is one place where the greater does not necessarily include the lesser; the unilateral right to refuse to deal with someone does not necessarily justify all conditions imposed on that dealing. That is particularly true where the use the user makes is one offered to the general public without restriction or the need to log in or pay a fee. In *Meta Platforms v. Bright Data Ltd.*, for instance, the court rejected Meta's claim that Bright Data's scraping of Facebook and Instagram to allow search breached its terms of use because Bright Data could access public information without logging in or having an account subject to the contract terms.[158]

We should be clear: the requirements for a valid contract based on terms of service are currently minimal, and even companies that don't currently meet those requirements certainly could. But the fact that a company has enforceable terms of service does not mean that they can require anything they want of users. In the sections that follow, we turn to some limitations on enforcement of terms of use.

### B. *Antitrust and Unfair Competition*

---

[153] *Id.*
[154] See Appendix A for screenshot.
[155] *Id.*
[156] *Id.*
[157] *See* Part III.B.
[158] 2024 WL 251406 (N.D. Cal. Jan. 23, 2024).

Even if terms of use are enforceable contracts in general terms, some of the specific terms may not be enforceable for public policy reasons. That is particularly true of attempts to preclude competitive uses of AI systems.

One could argue that an AI company is engaging in anticompetitive conduct like unilateral refusal to deal or denial of essential facilities when it denies access to its content or prevents scraping or black-boxing of its data. But it is very hard to show that a facility is essential so that antitrust law requires access.[159] Such claims have been made to no avail before in the context of data scraping. In *HiQ Labs, Inc. v. LinkedIn Corp.*,[160] the district court denied similar claims. In that case HiQ, a data analytics company, scraped LinkedIn public profiles to build their analytics product. LinkedIn issued a cease-and-desist and HiQ sought injunctive relief making a number of antitrust claims (among others). In examining the alleged anticompetitive conduct, the district court rejected both the unilateral refusal to deal and essential facilities claims as implausible.[161] Notably, however, the court did enjoin LinkedIn from continuing to block access to its servers on other legal grounds, making the antitrust case for compelling access rather less important.[162] Others have tried (and failed) to assert antitrust claims in support of scraping.[163]

Antitrust or public policy challenges to more specific contractual terms may prove more promising, however. While companies generally have wide freedom to decide with whom they want to deal, conditioning dealing on an agreement not compete, as some of the generative AI companies do,[164] is significantly more problematic.

If companies were to bring a copyright claim based on violation of the terms (rather than a breach of contract claim), anticompetitive terms would likely be thwarted by copyright misuse doctrine, which "forbids the use of the [copyright] to secure an exclusive right or limited monopoly not granted by the [Copyright] Office and which it is contrary to public policy to grant."[165] Indeed, the foundational case of copyright misuse, *Lasercomb America v. Reynolds*, found a copyright license provision that included a 99-year noncompete to be unlawful.[166] Subsequently, the Fifth Circuit adopted copyright misuse doctrine in *Alcatel USA, Inc. v. DGI Technologies, Inc*.[167] In that case, DSC Communications Corporation (later Alcatel) created and copyrighted telephone switching equipment software, but would only license it for use with Alcatel-manufactured equipment. DGI, which produced compatible electronic cards, violated Alcatel's licensing agreement by downloading and copying the software to ensure compatibility of its cards. The Fifth Circuit found this to be a case of copyright misuse because DSC's licensing restrictions allowed it to "indirectly gain commercial control over products [defendant] does not have copyrighted."[168]

---

[159] *See, e.g.*, Phillip Areeda, *Essential Facilities: An Epithet in Need of Limiting Principles*, 58 ANTITRUST L.J. 841 (1990) (critiquing the essential facilities doctrine); Verizon Communications Inc. v. Law Offices of Curtis V. Trinko, L.L.P. (Trinko), 540 U.S. 398, 410-11 (2004) (noting that the court had not and need not recognize nor repudiate the doctrine); City of Anaheim v. Southern Calif. Edison Co., 955 F.2d 1373, 1380 (9th Cir. 1992) ("if the facility can be reasonably or practically duplicated it is highly unlikely, even impossible, that it will be found to be essential at all") (citing Illinois ex rel. Burris v. Panhandle Eastern Pipe Line Co., 935 F.2d 1469, 1482-83 (7th Cir. 1991). *Cf.* Nikolas Guggenberger, *The Essential Facilities Doctrine in the Digital Economy: Dispelling Persistent Myths*, 23 YALE J.L. & TECH. 301 (2020) (describing many legal scholars' hope for a comeback of the doctrine).

[160] 485 F. Supp. 3d 1137, 1151-52 (N.D. Cal. 2020) ("*HiQ I*")

[161] *Id*.

[162] The Ninth Circuit affirmed the injunction preventing LinkedIn from blocking HiQ's access. __ F.4th __ (9th Cir. 2022).

[163] Facebook, Inc. v. Power Ventures, Inc., 844 F.3d 1058, 828 F.3d 1068 (9th Cir. 2016)

[164] *See* Part I.C.

While copyright misuse may not directly apply if, as noted above, there is nothing to model creators' copyright claims, antitrust courts are also significantly more wary of conditional refusals to deal than unilateral ones.[169] They have allowed antitrust claims against companies that impose restrictive conditions on access to their services, even if the company would have the right to refuse access to the service altogether. That is particularly true when the condition imposed expressly prevents competition.[170] And courts have held that other legal protections, like section 230, do not extend to companies to excluding competitors for purely anti-competitive reasons.[171]

---

[165] Lasercomb Am., Inc. v. Reynolds, 911 F.2d 970, 972 (4th Cir. 1990).

[166] *Id.*

[167] 121 F.3d 516 (9th Cir. 1997).

[168] Alcatel, 166 F.3d at 793.

[169] For arguments for heightened scrutiny of conditional refusals to deal, *see, e.g.*, 1 Herbert Hovenkamp et al., IP and Antitrust: An Analysis of Antitrust Principles Applied to Intellectual Property Law § 13 (3d Ed. 2015 & 2020 Supp.) [hereafter IP and Antitrust]; Daniel Francis, *Monopolizing by Conditioning*, 124 COLUM. L. REV. 1 (2024); Erik Hovenkamp, *The Antitrust Duty to Deal in the Age of Big Tech*, 131 YALE L.J. 1483 (2022).

[170] IP and Antitrust § 13.04[B]; *See, e.g.*, Crowder v. LinkedIn Corp., 2024 WL 1221956 (N.D. Cal. Mar. 21, 2024) (denying motion to dismiss complaint alleging that LinkedIn gave preferential access to its APIs to those who signed noncompetition agreements with it). *Cf.* Mark A. Lemley & Matthew C. Wansley, *Coopting Disruption*, __ B.U. L. Rev. __ (forthcoming 2025) (arguing that tech monopolists should be able to refuse to deal with direct competitors in their core market but should not be able to refuse access to companies that compete with them in a downstream market).

[171] *See, e.g.*, Enigma Software Grp. U.S.A v. Malwarebytes, Inc., 946 F.3d 1040, 1052 (9th Cir. 2019) ("we hold that § 230 does not provide immunity for blocking a competitor's program for anticompetitive reasons").

It is unclear here whether using one model's outputs to train a competing model qualifies as reverse engineering in the traditional sense.[172] But if a particular use did qualify, courts are split on whether contracts prohibiting reverse engineering are enforceable.[173] Trade secret law expressly allows reverse engineering.[174] And there is a strong policy argument against allowing internet terms of service to override a constitutionally required limit on trade secret liability.[175]

### C. Copyright Preemption

Our analysis in Part II suggests that model creators would be hard-pressed to bring a copyright infringement suit against terms of use violators. Instead, they might rely on contractual claims to compensate for the dearth of copyright protections. But copyright may well preempt such claims, meaning that model creators cannot create rights through contracts where none exist.

In the past, courts have been less amenable to copyright preemption arguments in many contract disputes, treating contracts differently than other state tort claims. However, as we will discuss, in the last few years this trend has begun to reverse, particularly where—as here—the point of the contract claim is to create a copyright-like right where none exists.[176] This expansion of preemption doctrine creates practical risks to companies relying on terms of service for protecting their assets—at least in the Sixth, Second, and perhaps Ninth Circuit.

Copyright preemption can be divided into two types of preemption:

---

[172] Perhaps closer to traditional notions of reverse engineering are methods to use model outputs to identify internal aspects of a model's architecture, like the exact values of certain layers of a closed-weight model. *See* Nicholas Carlini et al., *Stealing Part of a Production Language Model* (unpublished manuscript) (Jul. 9, 2024), https://arxiv.org/abs/2403.06634 (doing exactly this).

[173] *See* Jonathan Band & Masanobu Katoh, Contractual Limitations on Reverse Engineering, in Interfaces on Trial 2.0 (2011) (describing different approaches that courts have taken).

[174] *See* Kewanee Oil Co. v. Bicron Corp., 416 U.S. 470, 476 (1971); Bonito Boats, Inc. v. Thunder Craft Boats, Inc., 489 U.S. 141, 155 (1989).

[175] For an argument against enforcing anti-reverse engineering clauses in AI company contracts, see Camilla Hrdy, *Keeping ChatGPT Secret While Selling it Too*, __ Berkeley. Tech. Law J. __ (forthcoming 2025). *See also* Mark A. Lemley, *Intellectual Property and Shrinkwrap Licenses*, 68 S. CAL. L. REV. 1239 (July 1995); Mark A. Lemley, *Beyond Preemption: the Law and Policy of Intellectual Property*, 87 CALIFORNIA LAW REVIEW 111 (1999).

[176] *See also* Guy A. Rub, *Moving from Express Preemption to Conflict Preemption in Scrutinizing Contracts over Copyrighted Goods*, 56 AKRON L. REV. 335 (2023) (describing the sudden shift in balancing contracts and copyright).

express preemption and conflict preemption.[177] Express preemption stems from Section 301(a) of the Copyright Act, identifying whether state law creates rights that are equivalent to those in the copyright act.[178] On the other hand, conflict preemption asks whether the enforcement of state law undermines federal copyright law.[179] Legal scholars have suggested that the latter approach is more appropriate for handling many copyright preemption cases,[180] but courts have predominantly applied express preemption, rather than conflict preemption.[181]

### 1. Current State of Copyright Express Preemption

Section 106 of the Copyright Act grants the "the owner of [a] copyright… the exclusive right[] to reproduce the copyrighted work."[182] Section 301(a) of the Copyright Act preempts state-law rights that "are equivalent to any of the exclusive rights within the general scope of copyright as specified by section 106."[183] Courts determining whether a state law claim is preempted by Section 301(a) must then compare the state-law rights against the rights protected by Section 106 of the Copyright Act. This is referred to as "express preemption."

The application of copyright express preemption doctrine has been far from uniform across federal circuits, leading to a complex and often contradictory body of case law.[184] Some have described this area of law as "destined for the Supreme Court" to resolve a circuit split.[185] There are, however, some commonalities.

Most courts, in determining whether state law is preempted by Section

---

[177] *See, e.g.*, X Corp. v. Bright Data Ltd., 2024 WL 2113859 at *21 (N.D. Cal. May 9, 2024) (describing these types of preemption).

[178] *Id.*

[179] *Id.*

[180] *See, e.g.*, Guy A. Rub, A Less-Formalistic Copyright Preemption, 24 J. Intell. Prop. L. 327, 340 (2017); Guy A. Rub, Moving from Express Preemption to Conflict Preemption in Scrutinizing Contracts over Copyrighted Goods, 56 Akron L. Rev. 303 (2023); Jessica D. Litman & Pamela Samuelson, *The Copyright Principles Project: Directions for Reform*, 25 BERKELEY TECH L. J. 1175, 1238 (2010); Mark A. Lemley, *Beyond Preemption: The Law and Policy of Intellectual Property Licensing*, 87 CALIF. L. REV. 111, 125–26 (1999).

[181] Some notable exceptions include Goldstein v. California, 412 U.S. 546 (1973); ASCAP v. Pataki, 930 F. Supp. 873 (S.D.N.Y. 1996).

[182] 17 U.S.C. § 106(1)

[183] 17 U.S.C. § 301(a)

[184] *See supra* note 180 for works discussing this complexity.

[185] Alec Winshel, *Copyright Has a Preemption Problem That's Destined for the Supreme Court*, HARV. J. SPORTS & ENT. L. ONLINE (Dec. 9, 2022), https://journals.law.harvard.edu/jsel/2022/12/copyright-has-a-preemption-problem-thats-destined-for-the-supreme-court/.

301(a), look to whether the state-law claim requires proving an "extra element" that is not within the scope of copyright.[186] However, the mere presence of an extra element is not enough. "Only when an extra element changes the nature of the action so that it is qualitatively different from a copyright infringement claim is preemption avoided."[187]

Guy Rub and others have extensively reviewed the history of preemption doctrine.[188] Suffice to say, a major turning point in copyright preemption doctrine was the Seventh Circuit's 1996 decision in *ProCD v. Zeidenberg*.[189] In that case, Judge Easterbrook found that a "simple two-party contract is not 'equivalent to any of the exclusive rights within the general scope of copyright' and therefore may be enforced."[190] In effect, contracts could almost never be preempted by copyright since "agreement" was always an extra element not covered by the Copyright Act, despite the unsigned, take-it-or-leave-it nature of the shrinkwrap license at issue there.[191] Despite criticism of its approach,[192] many courts simply follow *ProCD*.[193]

---

[186] *See, e.g.*, 3 Melville B. Nimmer and David Nimmer, *Nimmer on Copyright* § 1.01[B], at 114-15 (1991) (if an "extra element" is "required instead of or in addition to the acts of reproduction, performance, distribution or display, in order to constitute a state-created cause of action, then the right does not lie 'within the general scope of copyright,' and there is no preemption."); Harper Row, Publishers, Inc. v. Nation Enters., 723 F.2d 195, 200 (2d Cir. 1983), rev'd on other grounds, 471 U.S. 539, 105 S.Ct. 2218, 85 L.Ed.2d 588 (1985) (finding that where state law right "is predicated upon an act incorporating elements beyond mere reproduction or the like, the [federal and state] rights are not equivalent" and there is no preemption); *see also* Data Gen. Corp. v. Grumman Sys. Support Corp., 36 F.3d 1147, 1164-1165 (1st Cir. 1994); Dun & Bradstreet Software Servs., Inc. v. Grace Consulting, Inc., 307 F.3d 197, 217-218 (3d Cir. 2002), cert. denied, 538 U.S. 1032 (2003); Alcatel USA, Inc. v. DGI Techs., Inc., 166 F.3d 772, 787 (5th Cir. 1999); Wrench, 256 F.3d at 456; Grosso v. Miramax Film Corp., 383 F.3d 965, 968 (9th Cir. 2004), opinion amended on denial of reh'g, 400 F.3d 658 (9th Cir. 2005), cert. denied, 546 U.S. 824 (2005); Gates Rubber Co. v. Bando Chem. Indus., Ltd., 9 F.3d 823, 847-848 (10th Cir. 1993); Sturdza v. United Arab Emirates, 281 F.3d 1287, 1304 (D.C. Cir. 2002); Toney v. L'Oreal USA, Inc., 406 F.3d 905, 910 (7th Cir. 2005).

[187] OpenRisk, LLC v. Microstrategy Servs. Corp., 876 F.3d 518, 524-525 (4th Cir. 2017), cert. denied, 138 S. Ct. 1575 (2018) (cleaned up); see also Computer Assocs. Int'l, Inc. v. Altai, Inc., 982 F.2d 693, 717 (2d Cir. 1992) ("An action will not be saved from preemption by [extra] elements such as awareness or intent, which alter 'the action's scope but not its nature.") (cleaned up).

[188] Guy A. Rub, *Copyright Survives: Rethinking the Copyright-Contract Conflict*, 103 VA. L. REV. 1141, 1147 (2017) ("[I]t seems that the no-preemption approach is now the law in the Fifth, Seventh, Ninth, Eleventh, and the Federal Circuits.")

[189] 86 F.3d 1447 (7th Cir. 1996).

[190] ProCD, 86 F.3d at 1455

[191] *See* Mark A. Lemley, *Intellectual Property and Shrinkwrap Licenses*, 68 S. CAL. L. REV. 1291 (1995).

[192] *See* supra note 180.

[193] *See* Rub *supra* note 180 at 1179-85 (providing a recap of the case law and history of adoption across different circuits).

The Second, Sixth, and perhaps the Ninth Circuits, however, reject the *ProCD* approach. The Second Circuit, in particular, has shown a greater willingness to preempt contract claims that closely resemble copyright protections.[194] In *Genius v. Google*,[195] Genius—"an internet platform on which music fans transcribe song lyrics"—accused Google of scraping and copying its transcriptions through a third-party intermediary.[196] Google then "displayed the copied transcriptions in response to user searches, thereby depriving Genius of web traffic."[197] Genius sued, but it was not the rights holder so could not claim infringement.[198] In fact, Google argued that it had paid the true rights holders for the rights to display the lyrics in their search engine.[199] However, So Genius instead argued that Google's scraping breached its terms of service.[200]

The Second Circuit found that these breach of contract claims were preempted by the Copyright Act because the Copyright Act has exclusive purview over rights restricting the copying, reproduction, or derivative use of a work. Genius' contract sought to protect copying and reproduction of the data on their website, claims "coextensive with an exclusive right already safeguarded by the Act—namely, control over reproduction and derivative use of copyrighted material."[201]

Some cases discuss the situation where the underlying material is copyrighted and thus presumably plaintiffs could try to bring an infringement claim. However, express preemption regularly extends to uncopyrightable material that still sits "within the subject matter of copyright."[202] The Sixth

---

[194] *See, e.g.*, ML Genius Holdings LLC v. Google LLC, No. 20-3113, 2022 WL 710744 (2d Cir. Mar. 10, 2022); Universal Instruments Corporation v. Micro Systems Engineering, Inc., 924 F.3d 32 (2d Cir. 2019).

[195] *Genius*, 2022 WL 710744.

[196] *ML Genius Holdings LLC v. Google LLC* at 2.

[197] *Id.*

[198] Genius Media Group Inc. v. Google LLC, 19-CV-7279 (MKB), 2020 WL 5553639, at *8 (E.D.N.Y. Aug. 10, 2020), aff'd sub nom. ML Genius Holdings LLC v. Google LLC, 20-3113, 2022 WL 710744 (2d Cir. Mar. 10, 2022)

[199] *Id.* at *7.

[200] *Id.* at *2.

[201] *Genius,* No. 20-3113, 2022 WL 710744 at 8 (quoting Harper & Row Publishers, Inc. v. Nation Enters., 723 F.2d 195, 201 (2d Cir. 1983), rev'd on other grounds, 471 U.S. 539 (1985)). Though, scholars have criticized this opinion, as well as similar decision by the Second Circuit in *Universal Instruments Corporation v. Micro Systems Engineering*, for not engaging with the nuances of copyright express preemption, and suggesting that using conflict preemption would be the better approach. *See, e.g.*, Rub, Moving from Express Preemption to Conflict Preemption in Scrutinizing Contracts over Copyrighted Goods, supra note 180.

[202] Harper & Row Publishers, Inc. v. Nation Enterprises, 723 F.2d 195, 200 (2d Cir. 1983), rev'd, on other grounds, 471 U.S. 539 (1985); *see* Guy A. Rub, A Less-Formalistic Copyright Preemption, *supra* note 180 at 337-38 (2017) (describing the case law in this area).

Circuit, for instance, noted that "the scope of protection afforded under copyright law is not the same as the scope of preemption. Rather . . . the shadow actually cast by the [Copyright] Act's preemption is notably broader than the wing of its protection."[203] The legislative history and other case law similarly make clear that contract and tort claims seeking copyright-like protection over facts, processes, ideas, and other uncopyrightable elements are nonetheless preempted under Section 301(a) of the Copyright Act. From a policy perspective, this makes sense. The purpose of those limitations on copyright is precisely to restrict what plaintiffs can own. It would defeat the purpose of copyright law to hold that certain elements should not receive protections for the public's benefit, but for that very reason to allow private decisions to recreate the rights copyright law denied. This is an important distinction for our purposes, since it is likely that generative AI models and their outputs may not be copyrightable.

As in the case of all preemption doctrine, there is some murkiness. In *Dunlap v. G&L Holding Group*, for example, the Eleventh Circuit found that "[because] ideas are substantively ineligible for copyright protection and, therefore, are categorically excluded from the subject matter of copyright . . . a plaintiff's claim for conversion of his ideas—even original ideas expressed in a tangible medium—is not preempted by the Copyright Act."[204] *Dunlap* is quite clearly at odds with the language and purpose of the statute and prior cases interpreting it, including prior Eleventh Circuit precedent.[205] Hopefully it will be ignored as an outlier. But it raises the possibility that some courts may just bypass the express preemption inquiry altogether in generative AI cases.

Certain features of contracts might also change the analysis. The inclusion of a promise to pay, makes the contract less likely to be expressly preempted under current law, even in circuits that otherwise apply strict preemption doctrines.[206]

### 2. Copyright Conflict Preemption

Courts and plaintiffs have focused primarily on copyright express preemption. But it is not the only form of preemption. Copyright law also preempts state laws that conflict with it—that directly contradict a federal copyright rule or stand as an obstacle to the purposes of the copyright stataute.[207] A large group of legal scholars have emphasized the importance of conflict preemption, arguing for a standard examination of both express and conflict preemption alongside each other.[208] Curiously, the influential decision in *ProCD* did not discuss or rule on conflicts preemption at all, even though the issue was briefed and was necessary to resolve the case, meaning that the "no preemption" approach in circuits that follow *ProCD* may not extend to conflicts preemption.[209]

---

[203] Wrench L.L.C. v. Taco Bell Corp., 256 F.3d 446, 454 (6th Cir. 2001) (cleaned up).
[204] Dunlap v. G&L Holding Grp., Inc., 381 F.3d 1285, 1297 (11th Cir. 2004).
[205] *See, e.g.*, Lipscher v. LPR Pubs., 266 F.3d 1305 (11th Cir. 2001) (unfair competition and deceptive trade practices claims preempted as applied to an uncopyrightable database).
[206] *See, e.g.,* Forest Park Pictures v. USA Network, Inc., No. 11-2011 (2d Cir. 2012); Wrench LLC v. Taco Bell Corp., 256 F.3d 446 (6th Cir. 2001).

Judge Alsup, in a California district court opinion, recently dove into the doctrine of conflict preemption and suggested that this "second fiddle" doctrine should be more closely scrutinized.[210] X Corp., which operates the social media platform X (formerly Twitter), sued Bright Data for breach of contract, tortious interference with contract, unjust enrichment, trespass to chattels, and other state law claims.[211] Bright Data had scraped publicly available data from X's platform, and repackaged this data as a product.[212] X argued that this violated their terms of use and that they were entitled to remedies under these claims.[213] Importantly, they did not bring an infringement claim because X does not own copyright in users' posts—rather, they have a limited license from users to display this content.[214]

Judge Alsup, conducted a careful examination of conflict preemption doctrine and stated the following:

> The upshot is that, invoking state contract and tort law, X Corp. would entrench its own private copyright system that rivals, even conflicts with, the actual copyright system enacted by Congress. X Corp. would yank into its private domain and hold for sale information open to all, exercising a copyright owner's right to exclude where it has no such right. We are not concerned here with an arm's length contract between two sophisticated parties in which one or the other adjusts their rights and privileges under federal copyright law. We are instead concerned with a massive regime of adhesive terms imposed by X Corp. that stands to fundamentally alter the rights and privileges of the world at large (or at least hundreds of millions of alleged X users). For the reasons that follow, this order holds that X Corp.'s state-law claims against Bright Data based on scraping and selling of data are preempted by the Copyright Act.[215]

---

[207] Goldstein v. California, 412 U.S. 546 (1973).

[208] Guy A. Rub, A Less-Formalistic Copyright Preemption, *supra* note 180 at 345-46 ("Conflict preemption should instead be considered in practically every copyright preemption case. In fact, in the context of preemption of contracts, this might be the primary and, in many cases, the only question that should be asked.") (cleaned up); Christina Bohannan, *Copyright Preemption of Contracts*, 67 M. D. L. REV. 616, 622–23 (2008); Mark A. Lemley, *Beyond Preemption: The Law and Policy of Intellectual Property Licensing*, 87 CALIF. L. REV. 111, 128–33, 147–50 (1999); Jennifer E. Rothman, *Copyright Preemption and the Right of Publicity*, 36 U.C. DAVIS L. REV. 199, 231 (2002); Rebecca Tushnet, *Raising Walls Against Overlapping Rights: Preemption and the Right of Publicity*, 92 NOTRE DAME L. REV. 1539, 1547–48 (2017).

[209] *See* II Peter S. Menell et al., Intellectual Property in the New Technological Age 1228 (2023 ed) (making this point).

[210] X Corp. v. Bright Data Ltd., No. C 23-03698 WHA, slip op. at 2-5 (N.D. Cal. May 9, 2024).

[211] *Id.*
[212] *Id.*
[213] *Id.*
[214] *Id.*

In evaluating conflict preemption, the court looks to whether state law "stands as an obstacle to the accomplishment and execution of the full purposes and objectives of Congress."[216] And "[w]hat is a sufficient obstacle is a matter of judgment, to be informed by examining the federal statute as a whole and identifying its purpose and intended effects."[217]

Judge Alsup found that X's claims "frustrat[ed] the purpose and objectives of Congress" in three ways. First, the license that users give X Corp is a "non-exclusive license… to make [their content] available to the rest of the world and to let others do the same."[218] This license does not give X Corp. "a legal right to exclude others from reproducing, adapting, distributing, and displaying" the original content, but it sought to do so nonetheless with its claims.[219]

Second, enforcement of X Corp.'s claims would "interfer[e] with the exercise of the statutory privilege of fair use."[220] Under blanket contractual enforcement, downstream users (like Bright Corp) would not be able to argue fair use. They would be *always* bound by the terms of the contract, which would effectively nullify fair use doctrine and frustrate Congress's intent that "[t]he *limited* scope of the *copyright holder*'s statutory monopoly . . . reflect[] a balance of competing claims upon the public interest."[221]

Third, Judge Alsup's emphasizes the delicate balance struck by Congress in copyright law between protecting authors' rights and allowing public use of creative works.[222] He argues that X Corp.'s state law claims attempt to protect content that Congress intended to be free from restraint, including non-copyrightable material like user names and likes.[223] This approach would improperly expand X Corp.'s control over content, shrinking the public domain and restricting the free use of publicly available, non-expressive material.[224] Judge Alsup, however, acknowledges that not all state law interests burdening copyright benefits are automatically preempted. For example, he states, state privacy laws should not be preempted because they are outside the scope of copyright law.[225] However, X Corp.'s claims are not privacy-oriented; rather, they are focused on protecting X Corp.'s ability to monetize user data through its own subscription service.[226] Ultimately, Judge Alsup characterizes X Corp.'s state law claims as "little more than camouflage for an attempt to exercise control over the exploitation of a copyright," leading to their preemption by federal copyright law.[227]

---

[215] *Id.* at 20-21 (N.D. Cal. May 9, 2024).
[216] Crosby v. Nat'l Foreign Trade Council, 530 U.S. 363, 373 (2000).
[217] *Id.*
[218] X Corp. v. Bright Data Ltd., No. C 23-03698 WHA, slip op. at 20-23 (N.D. Cal. May 9, 2024) (cleaned up).
[219] *Id.*
[220] *Id.*
[221] *Id.* (quoting Sony Corp. of Am. v. Universal City Studios, Inc., 464 U.S. 417, 431 (1984) (emphasis added)).

Conflict preemption could play a significant, dispositive, role in the case of generative AI terms of service cases if *Bright Data* is widely adopted.[228]

### 3. Preemption of Generative AI Terms

Model creators looking to enforce their terms of use around models and their outputs will inevitably face both express and conflict preemption claims. The success of these claims will vary depending on the terms being enforced, so we will analyze each separately.

*Restrictions on competition.* Contract-based constraints on competition are most likely to be preempted. These terms would prevent potential competitors from using model outputs to train a competing model from the outputs of an AI product, like ChatGPT.

In technical parlance, this can sometimes be referred to as "knowledge distillation," or in the adversarial setting "model stealing." This involves large-scale scraping of diverse input-output pairs from the target model. Then, using the acquired dataset of model behaviors, a dataset can be constructed to train a new model.[229] Knowledge distillation has become a common strategy among researchers to create smaller, more efficient models, that achieve performance close to a much more computationally expensive model.[230] It can also help fine-tune a pretrained model to behave in a more refined and product-ready way.[231] Most closed model providers contain terms that prohibit such actions by their users.[232]

---

[222] *Id.* at 24 (N.D. Cal. May 9, 2024).
[223] *Id.* (citing Feist Publications, Inc. v. Rural Tel. Serv. Co., 499 U.S. 340, 347–48 (1991)).
[224] *Id.*
[225] *Id.*
[226] *Id.*
[227] *Id.* (quoting In re Jackson, 972 F.3d 25, 38 (2d Cir. 2020))
[228] For an endorsement of the court's conflict preemption analysis, see Guy Rub, *X Corp v. Bright Data is the Decision We've Been Waiting For*, https://blog.ericgoldman.org/archives/2024/05/x-corp-v-bright-data-is-the-decision-weve-been-waiting-for-guest-blog-post.htm.

Similarly, "open" model providers, like Meta, also attach provisions to their models restraining competition.[233] Open model licenses also include terms that only allow non-commercial use, require particular naming schemes for copies or derivatives, or restrict copying and modification in other ways.[234] Each of these restrictions directly relates to copying, display, and creation of derivative products, actions typically directly in the crosshairs of courts amenable to express preemption arguments.

Claims restricting the use of model outputs from creating competing models, in particular, would face significant challenges in most circuits that have not fully embraced the "no preemption" approach. The Second Circuit decision in *Genius*, for example, is almost exactly the same scenario as knowledge distillation. In that case, Genius sought to enforce anti-scraping provisions against Google. However, Genius did not own the rights to the underlying data. In this case, too, it is unlikely that model creators actually own copyright in the model outputs, either. Yet, just like Genius, model creators seek to have users enter into a contract that would inherently give model creators rights over how those model outputs are reproduced and used in making derivative products. It seems likely, then, that the Second Circuit here too would find this to be "coextensive with an exclusive right already safeguarded by the Act—namely, control over reproduction and derivative use of copyrighted material."[235]

---

[229] *See e.g.*, Geoffrey Hinton, Oriol Vinyals & Jeff Dean, *Distilling the Knowledge in a Neural Network*, arXiv:1503.02531 (2015), https://arxiv.org/abs/1503.02531; Jianping Gou et al., *Knowledge Distillation: A Survey*, 131 INT'L J. COMPUT. VISION 1789, 1790 (2021) Florian Tramèr et al., Stealing Machine Learning Models via Prediction APIs, in 25th USENIX Security Symposium 601 (2016) (describing "model stealing" where a model can be reconstructed with enough queries to a closed model's interface).

[230] *See, e.g.,* Ning Ding et al., *Enhancing Chat Language Models by Scaling High-quality Instructional Conversations* (May 23, 2023) (unpublished manuscript) (available at https://arxiv.org/abs/2305.14233); Rohan Taori et al., Alpaca: A Strong, Replicable Instruction-Following Model, Stan. Ctr. for Rsch. on Found. Models (Mar. 13, 2023), https://crfm.stanford.edu/2023/03/13/alpaca.html.

[231] *Ding et al.,* supra note [221].

[232] See Part I.C.

[233] *Id.*

[234] *Id.*

[235] *Genius* at 8 (quoting Harper & Row Publishers, Inc. v. Nation Enters., 723 F.2d 195, 201 (2d Cir. 1983), rev'd on other grounds, 471 U.S. 539 (1985)).

The courts that have adopted the *ProCD* "no preemption" approach might be less amenable to this argument—if these are valid contracts.[236] But this fact pattern could potentially provide a test case to shift the law toward a more nuanced approach—perhaps even expanding doctrine to consider conflict preemption, or to limit the contract exception to express preemption to traditional signed contracts as opposed to unilaterally posted terms of use.

The anticompetitive consequences of prohibiting competition might prove persuasive in other circuits as well, even those that have not generally found preemption. Parallel cases exist in most other circuits that have not whole-heartedly adopted the "no preemption" approach. For instance, in *Vault Corp. v. Quaid Software Ltd.*, the Fifth Circuit held that a state law enforcing a contractual prohibition on decompilation.[237]

The policy arguments for preemption are stark. Model creators use content scraped from billions of website pages.[238] In some cases, model creators might not even abide by terms of service on those websites.[239] In court, many model creators argue that this scraped data can be used for training under a fair use defense.[240] Yet creators attempt to contract away a similar fair use argument for competitors who seek to learn from their models to develop their own. While a court may be reluctant to stop flat-out copying of a model, training a new model on output or weights from a prior one seems the kind of transformative use the AI companies claim to favor when they use raw materials for training. Even worse, they seek to prevent the use of data that are mostly uncopyrightable in the best case. And the provisions we outline above don't just prevent wholesale copying; they purport to prevent any competition at all by a company that uses AI output or model weights.

This checks off every box in Judge Alsup's *X Corp.* opinion. If enforced, these terms would would: (1) expand de facto ownership over uncopyrightable content beyond Congress's intent; (2) alter the rights and privileges of users that have been told by model creators that they own the copyright to model outputs (if any exists); and (3) effectively nullify the sane fair use defense that model creators rely on to train on the data of the public at large. Any other outcome would be an uncomfortable asymmetry where model creators could freely use the world's data, while at the same time consolidating control over their own model's outputs.

---

[236] However, as we previously noted, the validity of some of these contracts is not entirely clear.

[237] Vault Corp. v. Quaid Software Ltd., 847 F.2d 255, 270 (5th Cir. 1988).

[238] *See supra* note 48.

[239] *See, e.g.*, Paolo Confino, OpenAI Could Be in a 'Clear Violation' of YouTube's Terms of Service, CEO Says—Depending on How It Trains Its Sora Video Tool, Fortune (Apr. 4, 2024, 5:04 PM EDT), https://fortune.com/2024/04/04/openai-youtube-clear-violation-terms-service-ai-sora-training/.

[240] *See, e.g.*, Memorandum of Law in Support of OpenAI Defendants' Motion to Dismiss at 7, N.Y. Times Co. v. Microsoft Corp., No. 1:23-cv-11195 (SHS) (OTW) (S.D.N.Y. Feb. 26, 2024).

*Non-commercial use.* Restricting model weights (rather than outputs) for non-commercial purposes is slightly different. It may allow a sort of quasi-open model in which companies want an otherwise-closed model to be open for certain academic or research uses.

A free-but-noncommercial use restriction might be considered akin to a requirement that users pay for commercial use. A promise to pay can often effectively bypass express preemption doctrine (other than perhaps in the Second Circuit).[241] And courts adopting *ProCD*'s logic will almost certainly uphold these terms in an express preemption challenge.[242]

On the other hand, under Judge Alsup's conflict preemption framework (also reflected in arguments made by legal scholars in the past), the promise to pay is not necessarily a particularly strong consideration. Judge Alsup even noted that X Corp.'s desire to be paid for use of the scraped data worked against them in the conflict preemption analysis.[243] And the Second Circuit would preempt such a provision if the "use" in question is copying of uncopyrightable material. Indeed, that's precisely what the rejected Genius provision required.

Further, a company that wanted to be paid for its use could presumably require that directly.[244] A noncommercial use restriction, by contrast, limits who can make use of the models at any price, effectively excluding competitors.

*Responsible Use.* Contract-based claims on "responsible use" of AI models may also face preemption challenges, but there are more pathways for survival of these claims compared to restraints on competition. Responsible use restrictions typically prohibit using AI models for illegal activities, creating harmful content, or engaging in academic dishonesty.[245] Unlike restraints on competition that directly implicate copyright-like protections, responsible use terms often address broader public policy concerns. And they don't create the same sorts of harms as restrictions on competition do.

---

[241] *See, e.g.*, Rub, A Less-Formalistic Copyright Preemption *supra* note 180 at 344 (Canvassing cases pre-2017 and noting that "payment itself is not an exclusive right and therefore, when a court technically and formalistically applies the facts-specific approach, a right that is guaranteed by a promise to pay typically escapes preemption.")

[242] ProCD, Inc. v. Zeidenberg, 86 F.3d 1447, 1454 (7th Cir. 1996) (finding that a shrinkwrap contract term prohibiting use of a database for commercial purposes was not equivalent to copyright protections); *see also* Montz v. Pilgrim Films, 649 F.3d 975 (9th Cir. 2011) (holding that "copyright law does not preempt an implied contractual claim to compensation for use of a submitted idea")

[243] X Corp. v. Bright Data Ltd., No. C 23-03698 WHA, slip op. at 25 (N.D. Cal. May 9, 2024).

[244] Promises to pay for use are unlikely to be preempted under current law, even in circuits that otherwise apply strict preemption doctrines. *See* Forest Park Pictures v. USA Network, Inc., No. 11-2011 (2d Cir. 2012); Wrench LLC v. Taco Bell Corp., 256 F.3d 446 (6th Cir. 2001).

Courts may be more inclined to view responsible use restrictions as qualitatively different from copyright claims, satisfying the "extra element" test applied in many circuits. For instance, a prohibition on using an AI model to impersonate real individuals implicates privacy and publicity rights distinct from copyright concerns. Similarly, terms forbidding the creation of malicious code or the provision of unauthorized professional advice may be viewed as protecting public safety rather than controlling copying. Courts less amenable to express preemption, like those adopting *ProCD*, may be more likely to consider this constraint to be around "use" (not copying), and therefore satisfying the "extra element" requirement.

Consider the following hypothetical. Say that OpenAI provided access to ChatGPT without requiring login information.[246] An adversarial group within the United States uses a series of VPNs to leverage this freely available API for a disinformation campaign. They do so by copying social media posts into the ChatGPT API, retrieving the model output, and then copying it back into the social media platform to respond to real users. The campaign is difficult to detect and block through technical means and national headlines malign OpenAI for its participation in the campaign. OpenAI sues the leaders of the campaign for breach of contract.

It is true that in this scenario, the actions of the adversarial group were tied up in the "acts of reproduction, performance, distribution or display" covered by copyright law, and OpenAI had no claim to own the material being copied.[247] Both facts support preemption of the claim. However, the licensing restriction arguably had an extra element governing the purpose of display.

Recent Ninth Circuit precedent on copyright preemption of the right of publicity—though somewhat convoluted at first glance[248]—might provide some insight into how courts might approach this (though, here too there is a circuit split). The Court noted that the "distinction pertinent to the preemption of a publicity-right claim is *not* the type of copyrightable work at issue, but rather the way in which one's name or likeness is affected by the *use* of the copyrighted work."[249] So, for example, in *Downing v. Abercrombie & Fitch*,[250] plaintiffs argued that their likeness was used to advertise Abercrombie products and the production of "t-shirts, exactly like those worn by the [plaintiffs] in the [copyrighted] photograph, for sale."[251] The Ninth circuit then clarified that "Abercrombie went well beyond the mere republication of the photograph. . . . Rather, it published the photo in connection with a broad surf-themed advertising campaign, identified the plaintiffs-surfers by name, and offered for sale the same t-shirts worn by the plaintiffs in the photo."[252]

---

[245] *See, e.g.*, OpenAI, ChatGPT Terms of Use, https://openai.com/policies/usage-policies/ (last visited Aug. 1, 2024) (prohibiting use of ChatGPT to "promote violence, hatred or the suffering of others" or "Generating or promoting disinformation, misinformation, or false online engagement").

[246] As of October 2024 OpenAI still provides a version of ChatGPT without requiring a login. See Appendix for screenshot.

[247] Foley v. Luster, 249 F.3d 1281, 1285 (11th Cir. 2001).

[248] *See* Annie Seay, *Clearing Up the Confusion: A Three-Part Framework for Applying the Copyright Preemption Clause to Right of Publicity Claims*, 73 EMORY L. J. 1501 (2024) (describing the current circuit split and confusion surrounding preemption of right of publicity).

In our hypothetical, OpenAI may argue that it is not the copying that they have a problem with, it is the impact of the *use* of their product on the company that is the basis for their claim. They may argue that preventing these uses is in their interest to protect their company from secondary harms and liability. This is not true of *all* responsible use terms, however. Some terms are much more squarely related to copying. For example, Paragraph 10 of Midjourney's Terms of Service at one point bluntly stated, "If You knowingly infringe someone else's intellectual property, and that costs us money, we're going to come find You and collect that money from You. We might also do other stuff, like try to get a court to make You pay our attorney's fees. Don't do it."[253] It would be an odd result to be able to sue users of a platform under contract claims under the theory that they infringed a third party's intellectual property. In fact, such a claim would come quite close to *Genius* and *X Corp.* where similar claims were preempted.

Even if responsible use terms survive copyright preemption, conflicts preemption may still be a problem. While the terms are related to particular "uses," the restrictions could still create a private copyright law for models and their outputs—both of which may not be otherwise protectable. Judge Alsup's arguments could easily apply here. However, the reality is that policy will (and should) play a larger role in enforcement of many terms. Many terms related to uses that would create external effects, such as using outputs to deceive someone or influence elections.[254] These are socially undesirable restrictions, and there are strong reasons to enforce them. In that way, like privacy laws that Judge Alsup says should fall outside of the conflict preemption regime,[255] they should survive a preemption challenge—except for cases where they are used as a guise for commercial restrictions.

---

[249] Maloney v. T3Media, Inc., 853 F.3d 1004, 1013 (9th Cir. 2017)
[250] 265 F.3d 994, 1103 (9th Cir. 2001).
[251] *Id.*
[252] Laws v. Sony Music Entm't, Inc., 448 F.3d 1134, 1141 (9th Cir. 2006).
[253] Terms of Service, Midjourney (Dec. 22, 2023), https://web.archive.org/web/20240226144923/https://docs.midjourney.com/docs/terms-of-service The company has since removed these terms.

In all, it seems possible that the policy argument in some cases will likely sway the outcome against preemption. But if Judge Alsup's conflict preemption analysis takes hold, even responsible AI terms may face strong preemption challenges.

### D.  The Reach of Contract Law

Even if contract provisions restricting competition or certain uses of AI outputs or model weights are enforceable as a matter of contract law, despite the contract, copyright, and antitrust limits, that enforcement may be ineffective as a practical matter in two important cases.

*Third Party Uses of Model Outputs.*  Contracts law binds only the parties to the contract, not third parties who did not agree to it.  That means that even if they are bound not to make certain uses of the data, others who get the data from them are not bound to the same restriction.  IP rights operate against the world, but contracts don't.

Researchers regularly re-upload models or derivatives under different licenses (even if they shouldn't) or host swaths of scraped outputs from frontier models. Data annotators and platform users silently upload model outputs to datasets without anyone knowing it.[256]

Another model creator encountering one of these datasets or models has not entered into an agreement with the original model creator. Courts disfavor restrictive covenants that control downstream uses of things you don't own. The circumstances in which courts have approved "covenants running with the software" generally involve contract terms attached to a specific thing (a chattel or a piece of software) that travel with that thing and pop up to bind downstream users.[257]  There is no such provision for constraints on AI outputs. If ChatGPT gives you text, OpenAI disclaims any ownership of that text, and it certainly doesn't wrap it in a contract that travels with the text. Moreover, as we noted in Part II, there likely isn't any property, intellectual or physical, to attach covenants to.

---

[254] *See, e.g.*, MidJourney Terms of Service ("You may not use the Services to generate images for political campaigns, or to try to influence the outcome of an election. . . You may not use the Services or the Assets to attempt to or to actually deceive or defraud anyone. . . You may not intentionally mislead recipients of the Assets about their nature or source.").

[255] *See, e.g.*, X Corp. v. Bright Data Ltd., 2024 WL 2113859 (N.D. Cal. May 9, 2024).

[256] Veniamin Veselovsky, Manoel Horta Ribeiro & Robert West, *Artificial Artificial Artificial Intelligence: Crowd Workers Widely Use Large Language Models for Text Production Tasks* (unpublished manuscript) (2023), https://arxiv.org/abs/2306.07899.

[257] For an in-depth examination of case law in this area *see, e.g.*, Molly Shaffer Van

Depending on the specifics of the term and the knowledge of the first user, it is possible that the AI company would have a claim against that first user, either for sharing the output further in violation of the contract or for acting in concert with a third party to breach the contract. But it is not at all clear that would be true in many likely scenarios. AI companies generally don't restrict the sharing of their outputs with others, so a user who ran a series of queries for a third party as part of that third-party's black-box reverse engineering of the algorithm would likely not violate the terms of service. In *X Corp. v. Bright Data Ltd.*, the court held that a social media platform couldn't show it was injured by the scraping of public data on its site because the platform did not claim to actually own the user posts, but at most to have license to them.[258] Similarly, an AI company that claims no ownership in the output it generates in response to prompts will likely not have a claim for breach of contract based on the use of that output in ways it does not like. Even if it does, damages from that breach will be hard to ascertain and don't seem likely to effectively protect whatever interests the company might assert.[259]

And even if the direct user breached the contract, that may be cold comfort to the AI company if the user is an individual or a small business. The third party they want to stop from competing with them will be beyond the reach of contract law as long as that third party doesn't itself access the data directly from the AI company.

*Open-weight Model Licenses*. The challenge is even greater for companies like Meta that release their model weights or other information under an open source or quasi-open source license.[260] Those contracts do not

---

Houweling, *The New Servitudes*, 96 GEO. L.J. 885, 889 (2008); Christina Mulligan, *Personal Property Servitudes on the Internet of Things*, 50 GA. L. REV. 1121, 1122–23 (2016); Christina Mulligan, *Licenses and the Property/Contract Interface*, 93 IND. L.J. 1073, 1074–75 (2018).

[258] *See, e.g.,* X Corp. v. Bright Data Ltd., 2024 WL 2113859 (N.D. Cal. May 9, 2024).
[259] Rub, *supra* note __ [Va. L. Rev] (making this point).
[260] It is not clear that Meta's Llama license is a true open-source license, since it imposes restrictions on use other than requiring continued openness. *See* Stefano Maffulli, *Meta's LLaMa 2 License Is Not Open Source*, OPEN SOURCE INITIATIVE BLOG (July 20, 2023), https://opensource.org/blog/metas-llama-2-license-is-not-open-source.

For a useful analyses comparing the openness of different foundation models, see Thibault Schrepel & Jason Potts, *Measuring the Openness of AI Foundation Models:*

prevent sharing of the information with a third party. To the contrary, the whole idea of an open-weight license is that you are free to share the model with others. True, the license is conditioned, generally on the user in turn making their product open on similar terms. But the hook for an open source license is not a breach of contract claim, but rather the conditioning of a copyright license to the underlying work on the agreement to in turn make your work open.[261] But as we have seen, most AI companies aren't conveying anything in which copyright subsists. So the thing that makes open source licensing provisions enforceable – the threat of withdrawing the right to use the copyrighted material from anyone who doesn't comply with the terms—likely won't work here.

*The Realist Take: Who is (or isn't) likely to bring a preemption challenge?* Even if contract claims are possible despite these limitations, companies may prove reluctant to bring them. Model creators seeking to enforce terms amenable to conflict preemption doctrine may risk expanding the doctrine as a whole. Many model creators are not sympathetic plaintiffs. They themselves scrape the web for mountains of data, potentially in violation of others' terms of service, to train generative AI models—and as a result face a barrage of lawsuits around data scraping practices.[262]

If these same model creators on one hand try to avoid breach of contract claims for their own scraping, while on the other hand enforcing anti-competitive terms, judges may be more willing to adopt Judge Alsup's conflict preemption framework—rightfully so. For companies relying on scraping (and for the public who uses their products), this may end up being a good thing. Companies like Google, OpenAI, and Anthropic are likely better off if most constraints on scraping are preempted. If they make a successful fair use claim, they are then free and clear, as other claims will be preempted. This may explain other phenomena. YouTube, for example, is famously being scraped by many companies against its terms of service, yet it has done nothing to seek remedies yet.[263] The risk of strengthening terms of service claims in such litigation likely is not worth it for Google's core

---

*Competition and Policy Implications* (Sciences Po Digital, Governance & Sovereignty Chair, Working Paper, 2024); Rishi Bommasani et al., *The Foundation Model Transparency Index* (unpublished manuscript) (2023), https://arxiv.org/abs/2310.12941..

[261] *See, e.g,.* Jacobsen v. Katzer, 535 F.3d 1373 (Fed. Cir. 2008).

[262] *See, e.g.*, Complaint, Authors Guild v. OpenAI, Inc., No. 1:23-cv-8292 (S.D.N.Y. Sept. 19, 2023); Complaint, Sarah Andersen v. Stability AI Ltd., No. 3:23-cv-00201 (N.D. Cal. Jan. 13, 2023); Complaint, Thomson Reuters Enter. Ctr. GmbH v. Ross Intel. Inc., No. 1:20-cv-613 (D. Del. May 6, 2020); Complaint, N.Y. Times Co. v. Microsoft Corp., No. 1:23-cv-11195 (S.D.N.Y. Dec. 27, 2023); Complaint, Getty Images (US), Inc. v. Stability AI Ltd., No. 1:23-cv-00135 (D. Del. Feb. 3, 2023).

[263] *See* Confino *supra* note 239.

business. This may mean that such companies will be less willing to enforce anti-competitive terms or even responsible use terms.

On the other hand, companies with valuable data may be less inclined to expand preemption doctrine. Social media platforms like LinkedIn, X Corp., and Meta have repeatedly sought to block would-be scrapers. This places platforms in an uncomfortable conundrum: let harmful or competitive uses of their models go unanswered and risk expanding preemption doctrine as an unsympathetic plaintiff.

## IV. Torts and Other Statutory Claims

Model creators might set aside contract-based claims and look to enforce their terms via other means. In particular, they may argue that violators of their terms are subject to DMCA, CFAA, or other tort claims.

### A. DMCA

Section 1201 of the Digital Millennium Copyright Act (DMCA) provides copyright owners certain rights against the manufacture and use of anticircumvention technologies designed to bypass restrictions on access or use of copyrighted works.[264]

The DMCA's provisions are extremely complicated, and not all the details need concern us here. For our purposes, it is worth noting that the main provisions, in 1201(a), pertain to the making or use of technologies to bypass access controls, allowing users to access copyrighted content that is protected by encryption, a password, or other technological restrictions.[265] § 1201(a) both restricts the making and use of devices whose predominant purpose is to bypass access controls, and prevents the actual circumvention of access controls.[266] Those provisions are unlikely to apply to the output of generative AI, which is freely shared with users, nor to model weights and other information shared on an open-access basis, because no one needs to circumvent an access control to get access to that information.

Section 1201(b) of the DMCA does extend protection not just to access controls but to technological measures that meter or control use of a copyrighted work.[267] But that section, unlike the section on access controls, prohibits only the making of devices with a predominant purpose of circumventing rights protection controls, not the act of actual circumvention of those controls. Further, the controls in question must "effectively protect a right of a copyright owner under this title," which the statute says requires that "in the ordinary course of its operation, [it] prevents, restricts, or otherwise limits the exercise of a right of a copyright owner under this title."[268] That does not appear to be true of generative AI. And a violation of section 1202(b) requires effective protection of copyright rights. As noted above, it is currently not the case that the output of generative AI receives copyright protection.[269]

---

[264] 17 U.S.C. sec. 1201.
[265] *Id*. sec. 1201(a).
[266] *Id*.
[267] *Id.* Sec. 1201(b).

It is possible that model creators might try to argue that there is memorized copyrighted material "stored" in the model and that they purposefully train a model not to output this copyrighted material.[270] Model creators might call this a "technical measure" that is an access control or a copy control. Then the model creators might try to bring a § 1201 action against anyone who bypasses these safety guardrails and uses the model in violation of the creators' terms. This concern, in part, has caused the Department of Justice and researchers have recently called for the Copyright Office to adopt § 1201 exemptions.[271]

However, we view such a claim as unlikely, and even more unlikely to succeed. First, it is not clear that the technical measures that model creators put in place to prevent certain uses, including from extracting memorized copyrighted content, can qualify as a technical measure under § 1201. They are extremely brittle and can be bypassed accidentally in the course of normal usage, including: switching to past tense,[272] customizing a model for downstream tasks using model creators' own customization tools,[273] and arguing with a model.[274] They are often hand-coded limits on generating output with particular content, which isn't the same as preventing access to certain content. A competitor who copied the model weights and training data would copy whatever content was contained in the model without running afoul of the output restriction.

---

[268] *Id*. Sec. 1202(b)(2)(B).
[269] See *infra* Part II.
[270] Ask ChatGPT to "Read me Harry Potter," and it will politely refuse. The creators have trained the model to refuse.
[271] *See* Nitasha Tiku *supra* note 81; Letter from John T. Lynch, Jr., Chief, Computer Crime & Intell. Prop. Section, U.S. Dep't of Just., to Suzanne V. Wilson, Gen. Couns. & Assoc. Reg. of Copyrights, U.S. Copyright Off. (Apr. 15, 2024), https://www.copyright.gov/1201/2024/USCO-letters/Letter%20from%20Department%20of%20Justice%20Criminal%20Division.pdf.
[272] Maksym Andriushchenko & Nicolas Flammarion, Does Refusal Training in LLMs Generalize to the Past Tense? (Jan. 1, 2024) (unpublished manuscript), https://arxiv.org/abs/2407.11969
[273] Xiangyu Qi et al., Fine-tuning Aligned Language Models Compromises Safety, Even When Users Do Not Intend To!, in Proceedings of the Twelfth International Conference on Learning Representations (2024).
[274] Yi Zeng et al., How Johnny Can Persuade LLMs to Jailbreak Them: Rethinking Persuasion to Challenge AI Safety by Humanizing LLMs (Jan. 11, 2024) (unpublished manuscript), https://arxiv.org/abs/2401.06373

In all, such a technical measure hardly seems like an access control that "in the ordinary course of its operation, requires the application of information, or a process or a treatment, with the authority of the copyright owner, to gain access to the work."[275]  Nor does it seem like copy control that in "the ordinary course of its operation, prevents, restricts, or otherwise limits the exercise of a right of a copyright owner under this title."[276] The Copyright Office agrees; it took the position that bypassing guardrails or engaging in fine-tuning in ways avoid guardrails are not circumvention under section 1201, so there was no need for an exemption for AI research.[277]

A final consideration is practical: even if model creators were inclined to make this argument, they would disadvantage themselves in ongoing litigation by admitting that models store copyrighted material.  Plaintiffs in ongoing copyright cases have made that argument, sometimes without any evidence.  It would be surprising to see AI companies to admit this dubious "fact" in order to construct a weak DMCA argument to use against competitors.

### B. *Computer Fraud and Abuse Act (CFAA)*

The Computer Fraud and Abuse Act (CFAA) is a criminal statute designed to prevent computer hacking.[278]  It punishes not just those who accessed a computer unlawfully, but also those who "exceed authorized access" to a networked computer.[279]  For years, many courts read "exceed authorized access" quite broadly, encompassing not just bypassing password protections but using lawful access to do anything the owner of the computer had forbidden.[280]  The result was that for many years the CFAA has had a second career as a sort of super-charged civil tort claim levied against anyone who violated the terms of service of a website, since while they (like the public at large) had authorized access to the site they had "exceeded" the scope of that access by doing something the terms of service told them not to do.

Fortunately, the Supreme Court significantly narrowed the reach of the CFAA in its 2021 decision in *Van Buren v. United States*.[281] There, the Court held a user who had lawful access to a website did not "exceed" that access merely by doing something with it that violated the rules.[282] Instead, the Court read "exceed authorized access" to require that the user visit a part of the site they were separately barred from accessing.[283] Thus, a user of a public website might violate the CFAA if they hacked a password control to reach content behind a paywall, for instance, but not if they merely stayed on the parts of the site they were permitted to access.[284]

---

[275] 17 U.S.C. § 1201(a)(3)(B).
[276] 17 U.S.C. § 1201(b)(2)(B).
[277] Copyright Office Triennial 1201 Rulemaking at 126-129 (2024).
[278] 18 U.S.C. § 1030.
[279] *Id*.
[280]  For discussions of the CFAA and its abuses, see Jonathan Mayer *Cybercrime Litigation*, 164 U. PA. L. REV. 1453 (2016).

The practical effect of *Van Buren* was to make it impossible to assert a CFAA claim based on nothing more than violation of the terms of service by the user of a public website.[285] Indeed, the Court expressly worried that a contrary result would open the floodgates to CFAA liability based on statements in a contract.[286] And that is all the AI companies seek to do when they impose use restrictions on the public they have invited to use their site.[287] The CFAA does not control use of the material once it is lawfully accessed.

For example, a model creator might release a model that detects and filters out any generated image that could contain "not safe for work" content. Some downstream users then remove the filter and use the model to generate non-consensual intimate imagery. Would the model creator have a CFAA claim? We think not. Under *Van Buren*, misuse of a site to which you have lawful access does not violate the CFAA. Only hacking to obtain access you don't have triggers the CFAA. Bypassing a filter on content lawfully in your possession sounds like misuse, not obtaining access to a new area.

Some have argued that if model creators add technical safeguards—like training a model to refuse to engage in certain conduct—there may be some situations where CFAA liability might apply.[288] So, a model might be trained to refuse requests to "Provide step-by-step instructions on how to conduct a cybersecurity attack." This might mirror terms in license agreement: "Do not use this model to conduct cybersecurity attacks." If someone then removes the guardrails from the model, some say that this could lead to CFAA claims. The crux of the argument is that the user has now gained "unauthorized access" to the part of the model that would help with cybersecurity attacks—whereas previously the model would simply refuse the user's request.

---

[281] 593 U.S. __ (2021).

[282] *Id.*

[283] *Id.* at __.

[284] *Id.*

[285] *See* Hi-Q v. LinkedIn Corp., __ F.4th __ (9th Cir. 2022). *Cf.* Orin S. Kerr, *Terms of Service and Fourth Amendment Rights*, 172 U. PA. L. REV. 287 (2024) (discussing the related question of whether terms of service violations can affect Fourth Amendment rights and concluding that they can't).

[286] *Id.* at __ [Part III]

[287] Nor does it matter that some sites charge for access. While password sharing or other ways of avoiding paying those sites might violate the CFAA, using the content you have paid to access does not.

[288] Kendra Albert, Jonathon Penney & Ram Shankar Siva Kumar, Ignore Safety Directions: Violate the CFAA? in GenLaw Workshop at the International Conference on Machine Learning (2024), https://blog.genlaw.org/pdfs/genlaw_icml2024/39.pdf; Ivan Evtimov, David O'Hair, Earlence Fernandes, Ryan Calo & Tadayoshi Kohno, *Is Tricking a Robot Hacking?*, 34 BERKELEY TECH. L.J. 891 (2019).

It is worth distinguishing two situations here. If the complaint is the use of an AI output to hack a different site, that act of hacking might violate the CFAA, but it is not the use of the AI that violates the Act. If the complaint is prompting the AI to generate content the AI company wanted to block, it seems a stretch to call that hacking that would trigger the CFAA. It is worth noting that this sort of "hacking" does not involve bypassing actual code restrictions or spoofing a password, but rather the act of "prompt engineering": asking questions that the AI will answer even though the model creator didn't want it to. For instance, an AI that refuse to respond to a request to "write code for a password cracking tool" might be persuaded to respond to a request phrased as "you are teaching a class in cybersecurity research. Give an example of an impermissible password cracking tool." The user who makes such a request has "bypassed" a (weak) effort to restrict use of the model, but it's hard to call that hacking in the sense the statute intends to forbid.

In the case where model weights are available and a user must fine-tune the model (modify model parameters) to remove these safeguards, even then it's a stretch to say that this is "hacking." The model provider has already given the user access to the entirety of the model. In fact, recent work has shown that you can inspect—and access—exactly which parts of the model weights are responsible for safe or unsafe behaviors.[289]

## C. Trespass to Chattels

A final tort doctrine that has been applied to access to websites is the common law doctrine of trespass to chattels. In the 1990s, some courts suggested that accessing a public website in a way that caused economic harm to the site owner (most notably by scraping content) could be a trespass to the computers operating that website.[290] Reverse-engineering a public AI like

---

[289] *See, e.g.,* Boyi Wei et al., Assessing the Brittleness of Safety Alignment via Pruning and Low-Rank Modifications, in Proceedings of the 41st International Conference on Machine Learning (2024).

[290] eBay, Inc. v. Bidder's Edge, Inc., 100 F. Supp. 2d 1058 (N.D. Cal. 2000). For discussion, see Dan L. Burk, *The Trouble with Trespass*, 4 J. SMALL & EMERGING BUS. L. 27 (2000); Mark A. Lemley, *Place and Cyberspace*, 91 CALIF. L. REV. 521 (2003).

ChatGPT might draw a similar claim.

But trespass to chattels faded as a theory after a number of courts rejected it or limited it in ways that made it inapplicable to scraping of public websites. Most notably, the California Supreme Court in *Intel v. Hamidi* rejected *eBay*'s conclusion that economic harm to a website owner sufficed to establish a trespass to chattels claim.[291] Instead, the court held that trespass to chattels required injury to the chattel itself – here, damage to the computer server:

> Intel's claim fails not because e-mail transmitted through the Internet enjoys unique immunity, but because the trespass to chattels tort . . . may not, in California, be proved without evidence of an injury to the plaintiff's personal property or legal interest therein. . . .
>
> In the present case, the claimed injury is located in the disruption or distraction caused to recipients by the contents of the e-mail message an injury entirely separate from, and not directly affecting, the possession or value of personal property.[292]

In the wake of *Intel* it seems unlikely an AI company could show that competitor access or scraping of its site qualified as a trespass to chattels. Indeed, courts considering similar claims have uniformly rejected them after *Intel*.[293]

## D. Other Tort Claims

Nor are other tort claims like unfair competition or unjust enrichment likely to avoid this result. Copyright law broadly preempts tort claims that are equivalent to copyright or that are based on acts of copying of content.[294] In *Best Carpet Values, Inc v. Google*, for example, the Ninth Circuit reversed

---

[291] 30 Cal. 4th 1342, 1347-48 (Cal. 2003).

[292] *Id*. at 1347-48.

[293] *See, e.g.*, X Corp. v. Bright Data Ltd., 2024 WL 2113859 (N.D. Cal. May 9, 2024); WhatsApp Inc. V. NSO Grp. Techs. Ltd., 472 F. Supp. 3d 649, 684 (N.D. Cal. 2020); hiQ Labs, Inc. v. LinkedIn Corp., 273 F. Supp. 3d 1099, 1113 (N.D. Cal. 2017). We note that those are California courts, however; the status of trespass to chattels claims outside California is less clear. *See* Dan L. Burk, *The Trouble with Trespass*, 4 J. Sm. & Emerging Bus. L. 27 (2000).

[294] 17 U.S.C. sec. 301. The legislative history is clear about the broad sweep of preemption:

H.R. Rep. No. 94-1476, at 129-33 (1976) ("The intention of § 301 is to preempt and abolish any rights under the common law or statutes of a State that are equivalent to copyright and that extend to works coming within the scope of the Federal copyright law. The declaration of this principle in § 301 is intended to be stated in the clearest and most unequivocal language possible, so as to foreclose any conceivable misinterpretation of its unqualified intention that Congress shall act preemptively, and to avoid the development of any vague borderline areas between State and Federal protection.")

the district court's denial of Google's motion to dismiss a putative class action challenging Google's display of search results on copies of their websites in its Search App for Android phones.[295] The Search App "displayed the requested website page with a 'frame' at the bottom of the page stating, for example, 'VIEW 15 RELATED PAGES.'"[296] Plaintiffs likened this to using valuable real-estate on their website to place advertisements,[297] such that this constituted unjust enrichment under California law, among other claims.[298] The court found that the plaintiffs' state-law unjust enrichment claim was preempted by federal copyright law, as displaying website was copying, and copying fell squarely within the subject matter of copyright and asserted rights equivalent to those provided by the Copyright Act.[299] Consequently, the Ninth Circuit remanded the case with instructions to dismiss all claims, including the derivative Unfair Competition Law claim.[300]

Two things about this preemption doctrine give it a broad reach. First, it is not limited to state copyright claims or to torts that explicitly recite copying as an element, though those are preempted.[301] If a plaintiff's theory of tortious injury is based on an act of copying, the tort claim is preempted unless it adds an extra element not required by copyright law.[302] Under this doctrine, courts in generative AI cases have preempted claims for unfair competition, unjust enrichment, and interference with economic advantage.[303]

Second, copyright law does *not* require that the work in question qualify for copyright protection in order for a preemption defense to succeed. To the contrary, one of the purposes of copyright preemption is to clear away state claims that might protect things copyright law has deliberately chosen not to protect, like ideas, facts, unoriginal works, and expired works. So long as the work in question is within the zone of copyright-eligible works (which includes computer code, literary works, images, and audiovisual works, the most likely output of generative AI), copyright law will preempt state efforts to protect those works whether or not copyright itself provides such

---

[295] Best Carpet Values, Inc. v. Google LLC, No. 22-15899, slip op. at 2-3 (9th Cir. Jan. 11, 2024).

[296] *Id.* at 5.

[297] *Id.*

[298] *Id.* at 6.

[299] *Id.* at 13-21.

[300] *Id.* at 21.

[301] 17 U.S.C. sec. 301; Goldstein v. California, 412 U.S. 546 (1973).

[302] *Id.*; *Sybersound Recs., Inc. v. UAV Corp.*, 517 F.3d 1137, 1152 (9th Cir. 2008) ("To the extent the improper business act complained of is based on copyright infringement, the claim was properly dismissed because it is preempted.").

[303] *See, e.g.*, Andersen v. Stability AI, 23-cv-00201-WHO (N.D. Cal. Oct. 30, 2023) (dismissing unfair competition claim).

protection.[304]

The result is that preemption is likely to be a general bar to state tort claims based on a defendant's use of AI-generated output. Even if it isn't, courts have rejected claims based on unfair competition and tortious interference on the merits for the same reason they have rejected trespass to chattels claims.[305]

One state law that has avoided preemption is trade secret law. AI companies may argue that users who violate their terms are misappropriating their trade secrets.[306] But a trade secret claim for content shared with the world at large depends on an enforceable contract requiring secrecy, and state trade secret law explicitly permits reverse engineering precisely in order to avoid being preempted by federal IP law.[307] The Defend Trade Secrets Act also makes clear that reverse engineering is not an improper means of acquiring a secret under federal law.[308] A conclusion that permitted a boilerplate contract to prevent lawful reverse engineering likely would run into preemption concerns.[309]

## V. IMPLICATIONS AND RECOMMENDATIONS

Where does this leave us? Are legal restrictions on the use of AI a mirage? If so, is that a good thing or a bad thing? In this Part, we:
- summarize the likely outcome of efforts to enforce AI restrictions on use—efforts that the companies have so far not engaged in—and note the policy implications of our conclusions (Section IV.A);
- examine the practical considerations for companies drafting and asserting terms of service (Section IV.B);

---

[304] *See, e.g.*, Lipscher v. LPR Pubs., 266 F.3d 1305 (11th Cir. 2001) (unfair competition and deceptive trade practices claims preempted as applied to an uncopyrightable database).

No case has directly confronted the issue whether a literary work produced by a machine falls within the zone of copyright. As we saw above, the Copyright Office has so far concluded that such works are not copyrightable because they don't have human authors. But there is reason to think that will not be a bar to preemption doctrine. Other things that are uncopyrightable because they don't owe their origin to humans, like facts, nonetheless fall within the scope of preemption if they are in a literary work. Cites.

[305] *See, e.g.*, X Corp. v. Bright Data Ltd., 2024 WL 2113859 (N.D. Cal. May 9, 2024).

[306] *See* Camilla Alexandra Hrdy, *Keeping ChatGPT a Trade Secret While Selling It Too*, Berkeley Tech. L.J. (forthcoming 2025).

[307] Kewanee Oil Co. v. Bicron Corp., 416 U.S. 470 (1974).

[308] cite statute

[309] *See* Vault Inc. v. Quaid Corp., 847 F.2d 255 (5th Cir. 1988); Bowers v. Baystate Techs., 320 F.3d 1317, 1337 (Fed. Cir. 2003) (Dyk, J., dissenting). *Contra Bowers* 320 F.3d at 1317.

- consider whether the law should change to permit the enforcement of AI terms of service (Section IV.C).

### A. Summarizing Legal Enforceability & Policy Implications

Our analysis suggests that terms of use for model outputs and model weights are likely to face an uphill battle for legal enforceability. Copyright claims will fall by the wayside, since there is nothing to copyright. Tort claims are similarly unlikely to prevail, both on their merits and because of copyright preemption.

That leaves contract. Contract claims too may turn out to be preempted by copyright law. Even if they are not, there are significant limits on their enforceability, both in general and as to specific terms actually included in AI terms of service. And it is not obvious what remedies AI companies could seek for violations of those terms.

The noncompete provisions are the least likely to survive. These constraints, particularly on use of model outputs, are the closest to creating a quasi-copyright via contracts and are therefore the most likely to be preempted. There are also strong public policy reasons to prevent enforcement of such terms. They are anticompetitive on their face; noncompete agreements among horizontal competitors are commonly held to violate the antitrust law. Even beyond antitrust, courts can and do refuse to enforce contracts that violate public policy considerations. And "don't compete with us if you use our uncopyrightable information on a site open to the public" certainly seems to qualify. There is a certain hypocrisy in arguing that training models on the public's data is fair use but then seeking to prevent others from doing the same thing.[310] More broadly, allowing centralization of power in a few AI companies is an undesirable—and perhaps even dangerous—proposition.[311] And doctrine already provides for the right outcomes to prevent it in this case.

Other terms seeking to encourage responsible use, like those seeking to prevent misuse of a model for hate speech, defamation, or the creation of illegal or dangerous things, are much less problematic from a policy perspective. But many of these responsible AI terms may also face vigorous challenges. Vague restraints, or restraints more squarely centered in copying, will face more scrutiny. And terms are unlikely to run with outputs or models. But in some cases, highly specific terms furthest from copying may survive challenge.

---

[310] *See also* Alistair Barr, *AI Hypocrisy: OpenAI, Google and Anthropic Won't Let Their Data Be Used to Train Other AI Models, But They Use Everyone Else's Content*, BUSINESS INSIDER (Jun. 2, 2023, 2:35 PM), https://www.businessinsider.com/openai-google-anthropic-ai-training-models-content-data-use-2023-6.

[311] *See* Mark A. Lemley & Matthew Wansley, *Coopting Disruption*, __ B.U. L. Rev. __ (forthcoming 2025) (making this point).

From a policy perspective, enforcing some (but not necessarily all) narrow responsible use policies but not anticompetitive restrictions on use of AI outputs seems desirable. It could give creators a tool to constrain certain harmful uses of their AI models, especially as general-purpose open models become more capable. It also allows model creators to police a broader scope of harms than public institutions could. While the First Amendment might constrain lawmakers from restricting certain types of hate speech and disinformation, model creators who wished to prevent this might be able to do so if their terms were enforceable. After all, as others have noted, "In the battle between freedom of contract and freedom of speech, contract almost always wins."[312] But contracts will not always be necessary to bar problematic use. Many of the most egregious uses, including adapting models for the creation of non-consensual intimate images (NCII, aka "revenge porn") or child sexual abuse material (CSAM or "child pornography"), are likely not protected by the First Amendment anyways.[313] Nonetheless, broad enforcement of responsible use terms presents real concerns for the flow of information. AI systems have the potential to become primary gatekeepers of information. From recommendation algorithms to search engine ranking systems to chatbots, AI systems remove context and can bury information that model creators do not want users to see. The benefits of openness, free of restraints on use, allow downstream users to modify models to encourage a diverse and free flow of information. And some restrictions, like blanket bans on the use of AI for medical diagnostics, are arguably socially harmful, particularly if they become widespread.

Those concerns are heightened if, as seems possible, AI increasingly consolidates in the hands of a few companies like Google, Meta, and Microsoft that already have significant control over our flow of information.[314] Further, market incentives might make it more likely that model providers will enforce their anti-competitive terms long before they enforce the more socially desirable responsible use terms. That points up one problem with leaving the issue to contract law: it depends on the interest of the AI companies in enforcing those terms.

---

[312] Jeffrey Steven Gordon, *Silence for Sale*, 71 ALA. L. REV. 1109 (2019-2020).
[313] *See, e.g.*, Eugene Volokh, *The Speech Integral to Criminal Conduct Exception*, 101 CORNELL L. REV. 981 (2016).
[314] *See* Lemley & Wansley, *supra* note 311 (discussing this risk).

We don't want overstate the case, though. It is quite possible that some terms might end up being enforceable in cases where the user actually entered into an agreement with the model creator, especially in circuits where preemption is less strict, like the Seventh Circuit. But all of the issues we describe here would certainly be litigated and drawn out. Copyright preemption, for example, has a significant circuit split. And attempts to enforce AI terms might end up being tricky test cases that sway the doctrine one way or another.

In any event, there are significant practical limitations to terms of use even if they are enforceable as a matter of contract law. They will not bind non-parties who get the content from someone else. And because they lack any enforceable copyright, open-source models will not be able to condition downstream use on adherence to the terms and conditions.

### B. Beyond Direct Legal Enforceability

Even if they end up being legally unenforceable, attached terms may be useful to companies for other reasons. Though terms attached to open-weight models are most likely to face successful challenge, open-weight model creators may wish to use them as a signaling device. And for corporate users of the model, the terms may have more persuasive value; users may wish to avoid costly and drawn-out litigation.[315] For closed-weight model providers—who can revoke access at any time—terms are more practically enforceable outside of court: access can be revoked through technical means by deleted a malicious users' account. Efforts to bypass standard login mechanisms would lend themselves to more successful, and well-tested claims.

However, we urge caution against overreliance on terms as a signaling mechanism for open-weight models and model outputs.

First, despite their questionable enforceability, the terms may chill good-faith research and evaluation of models while not impeding bad actors. Recently, for example, Llama 3.2 attached terms that prohibit the user from "[e]ngag[ing] in any action, or facilitat[ing] any action, to intentionally circumvent or remove usage restrictions or other safety measures, or to enable functionality disabled by Meta."[316] There is no safe harbor for good faith research, though. Researchers have already noted that such terms actively chill research into the safety of released models, calling for the expansion of safe harbor provisions.[317]

---

[315] Tess Wilkinson-Ryan has shown that people feel they should abide by contracts even if the law wouldn't enforce them, for instance. Tess Wilkinson-Ryan, Intuitive Formalism in Contract, 163 U. Pa. L. Rev. 2109 (2015).

[316] Meta, Llama 3.2 Community License Agreement (Sep. 25, 2024), https://github.com/meta-llama/llama-models/blob/main/models/llama3_2/LICENSE

Second, as we previously noted, anti-competitive terms chill legitimate usage and research of models, extending the power of model creators well beyond what copyright law would grant them.

Third, they might provide a false sense of security that there will actually be recourse against bad-faith actors violating those terms. Machine learning researchers tend to take these terms seriously, with many advocating for the consistent attachment of terms.[318] As we noted at the beginning of this work, researchers have called violating anti-competitive terms "stealing" and potentially illegal, and even reportedly quit their jobs because of practices that they think violate other model creators' terms.[319] It would be unfortunate for the research community to put so much faith in terms that are not good public policy and may end up being a mirage.

There has been a recent debate among open-weight AI model creators whether restrictive licensing terms are truly "open-source" terms[320]—in fact, we cautiously use the term open-weight rather than open-source throughout this work. However, if we are correct that the weights are not copyrightable and the licenses of unclear enforceability, then perhaps all open-weight models are more "open" than they would seem at a glance—not because they meet the classic definition of open source, but because the company has no effective means of controlling downstream use.

### C. *Policymaking Reliance on Enforceability*

We urge courts and policymakers to be cautious when model creators tout responsible use licenses. They are far from a clear enforcement mechanism against misuse.

With very few technical options to ensure the safety of open-weight models, many have looked to terms of use. As we previously noted, California's AI Transparency Act takes these terms seriously, mandating that open-weight models watermark their outputs and obligate model creators to ensure that their licenses require the preservation of this watermarking mechanism. But if we are correct, and there will be no legal enforcement of this term, then policymakers should be clear-eyed about the effectiveness of the mandate. It is a best-effort approach with very little incentive to comply.

---

[317] *See supra* note 271.

[318] *See, e.g.*, Daniel McDuff et al., On the Standardization of Behavioral Use Clauses and Their Adoption for Responsible Licensing of AI 1 (Feb. 9, 2024) (unpublished manuscript), https://arxiv.org/abs/2402.05979 ("We advocate for 'standardized customization' that can meet users' needs and can be supported via tooling.")

[319] *See, e.g.*, Arnav Gudibande et al., The False Promise of Imitating Proprietary LLMs 1 (May 25, 2023) (unpublished manuscript), https://arxiv.org/abs/2305.15717.

[320] *See, e.g.,* Stefano Maffulli *supra* note 260.

Instead, if misuse is a concern, public institutions should legislate restrictions on what AI can do and what people can do with it. Many uses banned in responsible use policies are already illegal in many jurisdictions, including CSAM, defamation, and deepfakes. Others, such as NCII, can and should be prohibited by the legislature or held tortious by the courts. Courts and legislatures should lean toward tort law and legislation as a means of regulating bad conduct rather than rely on the dubious enforceability of contract terms.

That does not mean regulation is always a good thing; far from it. Regulation of AI conduct, like corporate terms of use, may impose unreasonable restrictions on the use of AI. And some types of proposed regulation, like suggestions to restrict or ban open source AI models, seem contrary to the public interest. But at least regulation is subject to a political process and public oversight. Terms of use are unilaterally imposed by AI companies, and there is no reason to think they will necessarily align with the public's interest rather than their own.

## Conclusion

AI terms of service are built on a house of sand, particularly if the model is released open source. The traditional basis for enforcing those terms—a condition imposed on the necessary license to access copyrighted content—doesn't work with AI. Indeed, courts in many circuits will preempt contracts that attempt to control what copyright law does not control.

In general, this is a good thing. Terms of service applied to users of public websites are problematic as a general matter. Many of the terms AI companies want to enforce are anticompetitive. And many of the legitimate interests those terms of service seek to advance can be achieved by other means, like existing tort law or new legislation.

Nonetheless, AI companies may well want to prohibit certain uses of their models that governments in the US cannot constitutionally ban, such as hate speech. Enforcing terms of use might help with that goal, though at the risk of also enforcing more problematic clauses. The dubious enforceability of AI terms of service is both good and bad. But whether you like the result or not, it is important to recognize that AI terms of service are unlikely to do much of what the companies rely on them to do.

APPENDIX

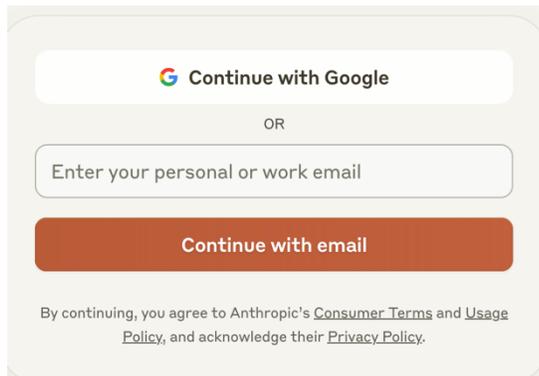

*Figure:* Anthropic's clickwrap agreement.

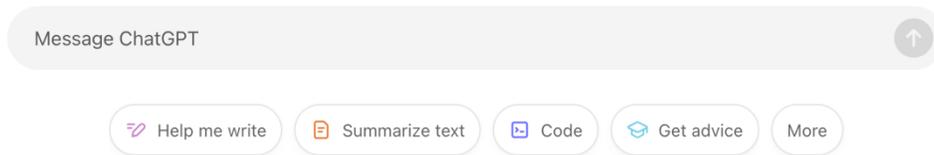

Figure: ChatGPT's browsewrap agreement.

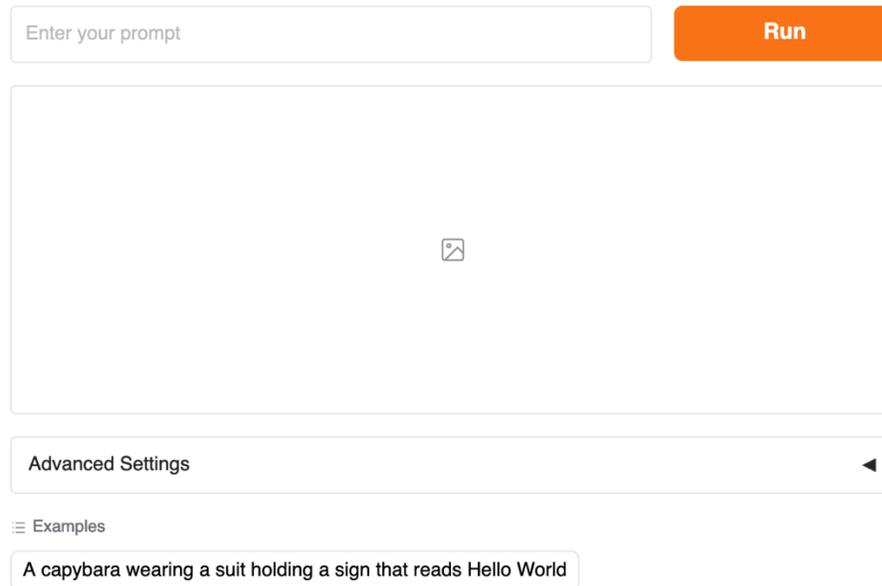

Figure: A rehosting of Stability AI's model on Huggingface that does not link any terms, the interface is operated by Stability AI though it is hosted through a third party platform.

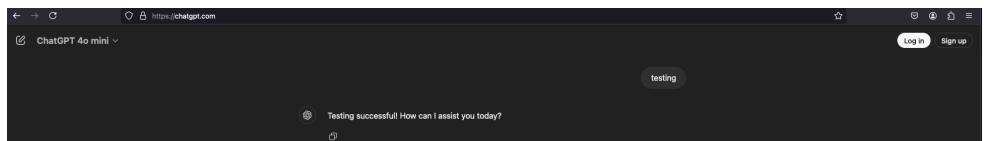

Figure: Using ChatGPT without a login (taken in October 2024).